\documentclass[twocolumn,showpacs,aps,prd,amsmath,amssymb]{revtex4-2}
\usepackage{graphicx}
\usepackage{dcolumn}
\usepackage{amsmath}
\usepackage{epsfig}
\usepackage{subfigure}
\usepackage{epstopdf}
\usepackage{multirow}
\usepackage{mathrsfs}
\usepackage[bookmarksnumbered, pdfstartview=FitH,colorlinks,urlcolor=blue, citecolor=blue,linkcolor=blue,] {hyperref}

\usepackage{amssymb} 
\usepackage{bm}
\usepackage{color}
 \usepackage{xspace}
\usepackage{textpos}
\usepackage[english]{babel}

\usepackage{overpic}
\usepackage{colortbl}
\usepackage{rotating}

\newcommand{\psip}{\psi(2S)}

\newcommand{\jpsi}{J/\psi}
\newcommand{\chicJ}{\chi_{cJ}}

\newcommand{\EE}{e^+e^-}
\newcommand{\ee}{e^+e^-}
\newcommand{\MM}{\mu^+\mu^-}
\newcommand{\pip}{\pi^+}
\newcommand{\pim}{\pi^-}
\newcommand{\piz}{\pi^0}
\newcommand{\pp}{\pi^+\pi^-}

\newcommand{\ppjpsi}{\pi^+\pi^-\jpsi}

\newcommand{\bfg}{\begin{figure}}
\newcommand{\efg}{\end{figure}}
\newcommand{\bitm}{\begin{itemize}}
\newcommand{\eitm}{\end{itemize}}
\newcommand{\bnum}{\begin{enumerate}}
\newcommand{\enum}{\end{enumerate}}
\newcommand{\btbl}{\begin{table}}
\newcommand{\etbl}{\end{table}}
\newcommand{\btbu}{\begin{tabular}}
\newcommand{\etbu}{\end{tabular}}

\newcommand{\GG}{\gamma\gamma}

\newcommand{\LL}{\ell^+\ell^-}
\newcommand{\beq}{\begin{equation}}
\newcommand{\edq}{\end{equation}}

\newcommand{\tx}{\tilde{X}(3872)}
 
\newcommand{\jpc}{J^{PC}}

\newcommand{\etap}{\eta^{\prime}}

\newcommand{\hc}{h_{c}}

\newcommand{\kk}{K^+K^-}
\newcommand{\pppsip}{\pi^+\pi^- \psip}

\newcommand{\eeq}{\end{equation}}

\newcommand{\mevcc}{\mathrm{MeV}/c^2}
\newcommand{\gev}{\mathrm{GeV}}

\newcommand{\gevcc}{\mathrm{GeV}/c^2}
\newcommand{\dz}{D^{0}}
\newcommand{\dsm}{D^{*-}}
\newcommand{\dm}{D^{-}}
\newcommand{\dsz}{D^{*0}}
\newcommand{\dplus}{D^{+}}

\usepackage{lineno}
\usepackage{enumerate}
\usepackage{float}
\usepackage{color}

\begin{document}

\title{\bf \boldmath
Cross section measurement  of $\EE\to \eta\psi(2S)$ and search for  $\ee\to\eta\tilde{X}(3872)$  
}

 \author{
 M.~Ablikim$^{1}$, M.~N.~Achasov$^{4,c}$, P.~Adlarson$^{75}$, O.~Afedulidis$^{3}$, X.~C.~Ai$^{80}$, R.~Aliberti$^{35}$, A.~Amoroso$^{74A,74C}$, Q.~An$^{71,58,a}$, Y.~Bai$^{57}$, O.~Bakina$^{36}$
% ...
\\
\vspace{0.2cm}
(BESIII Collaboration)\\
\vspace{0.2cm} {\it
$^{1}$ Institute of High Energy Physics, Beijing 100049, People's Republic of China\\
% ...
$^{3}$ Bochum  Ruhr-University, D-44780 Bochum, Germany\\
$^{4}$ Budker Institute of Nuclear Physics SB RAS (BINP), Novosibirsk 630090, Russia\\
% ...
$^{35}$ Johannes Gutenberg University of Mainz, Johann-Joachim-Becher-Weg 45, D-55099 Mainz, Germany\\
$^{36}$ Joint Institute for Nuclear Research, 141980 Dubna, Moscow region, Russia\\
% ...
$^{57}$ Southeast University, Nanjing 211100, People's Republic of China\\
$^{58}$ State Key Laboratory of Particle Detection and Electronics, Beijing 100049, Hefei 230026, People's Republic of China\\
% ...
$^{71}$ University of Science and Technology of China, Hefei 230026, People's Republic of China\\
% ...
$^{74}$ University of Turin and INFN, (A)University of Turin, I-10125, Turin, Italy; (B)University of Eastern Piedmont, I-15121, Alessandria, Italy; (C)INFN, I-10125, Turin, Italy\\
$^{75}$ Uppsala University, Box 516, SE-75120 Uppsala, Sweden\\
% ...
$^{80}$ Zhengzhou University, Zhengzhou 450001, People's Republic of China\\
\vspace{0.2cm}
$^{a}$ Deceased\\
% ...
$^{c}$ Also at the Novosibirsk State University, Novosibirsk, 630090, Russia\\
% ...
}
}

\date{\today}

\begin{abstract}
The energy-dependent cross section for $\EE\to \eta\psi(2S)$ is measured  at eighteen center of mass energies from 4.288 GeV to 4.951 GeV using the  BESIII detector. 
Using the same data samples, we also perform the first search for the reaction $\ee\to\eta\tilde{X}(3872)$, but no evidence is found for the $\tilde{X}(3872)$ in the   $\ppjpsi$ mass distribution.  
At each of the eighteen center of mass energies, upper limits at the 90\% confidence level  on the cross section for $\ee\to\eta\psi(2S)$ and   on the product of the $\ee\to\eta\tx$  cross section with the branching fraction of   $\tx\to\ppjpsi$  are reported.
\end{abstract}

\maketitle

\section{\boldmath INTRODUCTION}
\label{introduction}

Since the $X(3872)$ state was first observed by the  Belle experiment in 2003~\cite{x3872}, abundant   unexpected  charmonium-like states, such as the  $Y(4260)$, $Z_{\rm c}(3900)$, and $Z_{\rm cs}(3985)$ states,  have been discovered by the Belle, BaBar, BESIII, and CLEO experiments~\cite{ theory-Y-states-Brambilla-2020, zcs3985}. Their properties differ from conventional charmonia  and do not match predictions based on potential model calculations for the charmonium spectrum~\cite{chao2}.  Therefore, these states are
collectively known as  the ``$XYZ$'' particles and regarded as  
  exotic states.    They are still not  well understood and  have many theoretical interpretations, including compact tetraquarks, molecules,
hybrids, and  hadrocharmonia~\cite{theory-Y-states-chenhuaxing-2016, theory-Y-states-Esposito-2017, theory-Y-states-Richard-2017, theory-Y-states-Ali-2017, theory-Y-states-Stephen-2018,  theory-Y-states-guofenghun-2018, theory-Y-states-Brambilla-2020}, among others.
 
 Extensive experimental measurements of the production and decays of such states are essential for the understanding of their internal structures  and for  deepening our understanding of  the   low-energy properties  of the strong interaction.
 Over the past two decades, several vector  charmonium-like states,  including the $Y(4260)$~\cite{intro-BaBar-Y4260, intro-BaBar-Y4260-2012, intro-Belle-Y4260, intro-Belle-Y4260-2, intro-CLEO-Y4260-1},
 $Y(4360)$~\cite{intro-BaBar-Y4360, intro-BaBar-Y4360-Y4660-2014, intro-Belle-Y4360-Y4660},
and $Y(4660)$~\cite{intro-BaBar-Y4360-Y4660-2014, intro-Belle-Y4360-Y4660}, have been observed in the processes $\ee\to\ppjpsi$ and  $\pppsip$. 
In addition to   these two processes through $\pi\pi$ hadronic transitions, other hadronic transitions   of these $Y$ states to
lower mass charmonia  ($\jpsi,\psip$, $\chicJ(J=0,1,2)$, $\hc$, etc.)   provide further insight into their internal structures. 

Besides the decays to the above  mentioned hidden-charm final states, $Y$ states  have  also been discovered in the processes of $\ee$ annihilating to open-charm final states, such as $\ee\to Y(4230)\to\dz\dsm\pip+c.c.$~\cite{intro-Y4220-bes3-open-charm}, $\ee\to Y(4360)\to\dplus\dm\pip\pim$~\cite{intro-Y4360-bes3-open-charm-2022}, and $\ee\to  Y(4660)\to D_{s}^{+}D_{s1}(2536)^{-}$~\cite{Y4630-belle-2020}.  Very recently,  the $Y(4230)$, $Y(4500)$, and $Y(4660)$  states have been observed in the  process $\ee\to\dsz\dsm\pip$ for the first time by BESIII~\cite{Ystate-DDpi-bes3-2023},   where the $Y(4500)$ is compatible with the state observed in  $\ee\to\kk\jpsi$~\cite{kkjpsi-bes3-2022}.

Since the BESIII experiment observed the $Y(4230)$ in the $\ee\to\eta\jpsi$ and $\eta'\jpsi$ processes, it follows that the  process $\ee\to\eta\psip$ can be an important way  to search for  $Y$ states.  The CLEO-c experiment searched for this process using data at a single center of mass (c.m.) energy $\sqrt{s}=4.260$~GeV, but it did not observe a signal.  
 Using a total of $5.25~{\rm fb}^{-1}$ of $e^+e^-$ collision data with c.m.\ energies from 4.236 to 4.600 GeV, BESIII reported the first observation of the process $\EE\to \eta\psip$ with a statistical significance of 4.9 standard deviations~\cite{etapsip}. 
  In the past two years,  BESIII has collected more data at  c.m.\ energies above 4.600 GeV.
  In this work, we update the measurement for the cross sections of $\ee\to\eta\psi(2S)$ with the new  data samples. 
  
In addition, we search for the $\tilde{X}(3872)$ state reported by the COMPASS experiment~\cite{compass}.  
 This state has quantum numbers $\jpc=1^{+-}$ and a mass of 3860 MeV/$c^2$,  therefore, it is  regarded as the partner state of the $X(3872)$,  since their masses and widths are close to each other but with different $C$ parities. 
 Inspired by Ref.~\cite{compass},  
 we   search for the  $\tx$ in the   process  $\ee\to\eta\tx$, with    $\tilde{X}(3872)\to\ppjpsi$.

In this article,  we analyze the data   collected with the BESIII detector~\cite{Ablikim-2009aa} 
 at  c.m.\ energies from 4.288 GeV to 4.951 GeV; the corresponding c.m.\ energies and luminosities are listed in  Table~\ref{table:Preliminary results}. 
The energies were measured using $\EE\to \MM$ events with an uncertainty of 0.8~MeV~\cite{bes3-energy-measurement}, while the integrated luminosities were measured using Bhabha scattering events with an uncertainty of 1.0\%~\cite{luminosity-measurement, luminosity-measurement-2}. We reconstruct the $\ee\to\eta\psip/\tx$ signal process  using the
decays  $\psip/\tx\to\ppjpsi$, $\jpsi \to \LL$ ($\ell = e$ or $\mu$), and  $\eta\to\gamma\gamma$.

\section{\boldmath BESIII detector and Monte Carlo simulation}
\label{sec:detector}

The BESIII detector records symmetric $e^+e^-$ collisions provided by the BEPCII storage ring~\cite{Yu-IPAC2016-TUYA01} in the c.m.\ energy range from 2.00 to 4.95~${\rm GeV}$, with a peak luminosity of $1 \times 10^{33}\;\text{cm}^{-2}\text{s}^{-1}$
achieved at $\sqrt{s} = 3.77\;\text{GeV}$. 
BESIII has collected large data samples in this energy region~\cite{Ablikim:2019hff, EcmsMea, EventFilter}.
The cylindrical core of the BESIII detector consists of a helium-based multilayer drift chamber (MDC), a plastic scintillator time-of-flight system (TOF), and a CsI(Tl) electromagnetic calorimeter (EMC), which are all enclosed in a superconducting solenoidal magnet providing a 1.0~T magnetic field. The solenoid is supported by an octagonal flux-return yoke with resistive plate chamber muon identifier modules interleaved with steel. The acceptance of charged particles and photons is 93\% over $4\pi$ solid angle. The charged-particle momentum resolution at $1~{\rm GeV}/c$ is $0.5\%$, and the $dE/dx$ resolution is $6\%$ for the electrons from Bhabha scattering events. The EMC measures photon energies with a resolution of $2.5\%$ ($5\%$) at $1$~GeV in the barrel (end cap) region. The time resolution of the TOF barrel part is 68~ps.  The end cap TOF system was upgraded in 2015 with multi-gap resistive plate chamber technology, used for the data taking of this analysis, providing a time resolution of 60~ps~\cite{etof, etof-2, etof-3}.

Simulated data samples produced with a {\sc geant4}-based~\cite{geant4}  Monte Carlo (MC) package, which includes the geometric description of 
the BESIII detector and the detector response, are used to determine detection efficiencies and to estimate background contributions. 
The simulation models the beam energy spread, the initial state radiation (ISR), and the vacuum polarization in the $e^+e^-$ annihilations with 
the generator {\sc kkmc}~\cite{ref-kkmc, ref-kkmc-2}.
 
Signal MC samples for the processes $\EE\to \eta\psip$ and $\eta\tx$ are generated using helicity-amplitude ({\sc helamp})~\cite{ref-evtgen, ref-evtgen-2}  and phase-space ({\sc phsp})~\cite{ref-evtgen, ref-evtgen-2} models, respectively, with  200,000 events at each c.m.\ energy. 
The $\tx$ mass is set to   $3860.0\pm10.4~\mevcc$, from the COMPASS measurement~\cite{compass}. We study two different scenarios for the $\tx$ width.  In setting I, the width is set to 51 MeV based on the COMPASS result~\cite{compass}; and in setting II, it is set to 1.19 MeV based on the $X(3872)$ width quoted from the Particle Data Group (PDG)~\cite{pdg}. We provide the $\ee\to\eta\tx$ cross section under these two different  assumptions. 

An inclusive  MC sample  at $\sqrt{s}=4.682$ GeV is used to  estimate possible background contributions.  It is 40 times the size of the data sample
 and  includes the production of open-charm processes, the ISR production of vector charmonium(-like) states, and the continuum processes incorporated in {\sc kkmc}. 
The known decay modes are modelled with {\sc evtgen}~\cite{ref-evtgen, ref-evtgen-2} using branching fractions taken from the
PDG~\cite{pdg}, and the remaining unknown charmonium decays are modelled with {\sc lundcharm}~\cite{ref:lundcharm, ref:lundcharm2}. 
Final-state radiation~(FSR) from charged final-state particles is incorporated using  {\sc photos}~\cite{photos}.
Exclusive MC samples for some background processes  are also generated   at each c.m.\ energy to study their line shapes.  

 \section{\boldmath Analysis of $\ee\to\eta\psip$}
\label{sec:de1}

\subsection{Event selection}
\label{sec:eventselection1}
The criteria applied in this analysis to select charged tracks and photons are described in Ref.~\cite{etapsip}.
Candidate events are required to have four charged tracks with zero net charge and at least two photon candidates.  The pions and leptons have distinct momentum distributions for the signal process, thus the charged particles with momenta less than $0.8~\gev/c$ in the laboratory frame are assigned  to be $\pi^{\pm}$, whereas the ones with momenta greater than $1.0~\gev/c$ are assumed to be $\ell^{\pm}$. 
 The  EMC energy deposits of electron   and muon candidates are required to be greater than $1.0~\gev$ and less than $0.4~\gev$, respectively. Photon candidates are reconstructed from showers in the EMC crystals. The reconstructed energies for the clusters in the barrel ($|\!\cos{\rm \theta|}<0.80$) and the end caps ($0.86<|\!\cos{\rm \theta}|<0.92$) of the EMC are required to be higher than 25 MeV and 50~MeV, respectively.  To eliminate showers associated with    charged particles, the angle between the photon and any charged track   is required to be greater than 10 degrees.
To suppress the electronic noise and energy deposits unrelated to the event,  the time of the EMC shower is required to be $0\leqslant t \leqslant 700$~ns with respect to the event start time. 

To improve the mass resolution and suppress the
background,   charged tracks are required to originate from a common
vertex, and  a four-constraint (4C) kinematic fit imposing energy-momentum conservation  under the hypothesis of $\ee\to\gamma\gamma\pip\pim \ell^{+}\ell^{-}$ is performed. If there are more than two photons in an event, the combination of $\gamma\gamma\pip\pim \ell^{+}\ell^{-}$ with the smallest chi-square $\chi^{2}_{\rm 4C}$  is retained for further study. 
 The $\chi^{2}_{\rm 4C}$ of surviving events is required to be less than 40.  

To identify signal candidates that involve the $\jpsi$ resonance, the $\LL$ invariant mass is required to satisfy  $3064.6 < M(\LL)< 3140.8$~MeV/$c^2$, which is about three times the resolution of the $\jpsi$ nominal mass. To remove the background events $\EE\to\etap\jpsi$ with $\etap\to\pip\pim\eta$, the invariant mass of $\pip \pim \gamma\gamma$, $M(\pip \pim \gamma\gamma)$, is required to be greater than $1~\gevcc$.  
 At the energy points above 4.600 GeV, the invariant mass of  $\GG\jpsi$ is required to be greater than 3.74 $\gevcc$ to remove the backgrounds from the $\ee\to\pp\psip$ process.

\begin{figure*}[tbp]
\renewcommand\figurename{\rm FIG}
\centering
\subfigure{
\label{fig:2d-mass-eta-pipijpsi-a}
\includegraphics[width=0.25\paperwidth]
 {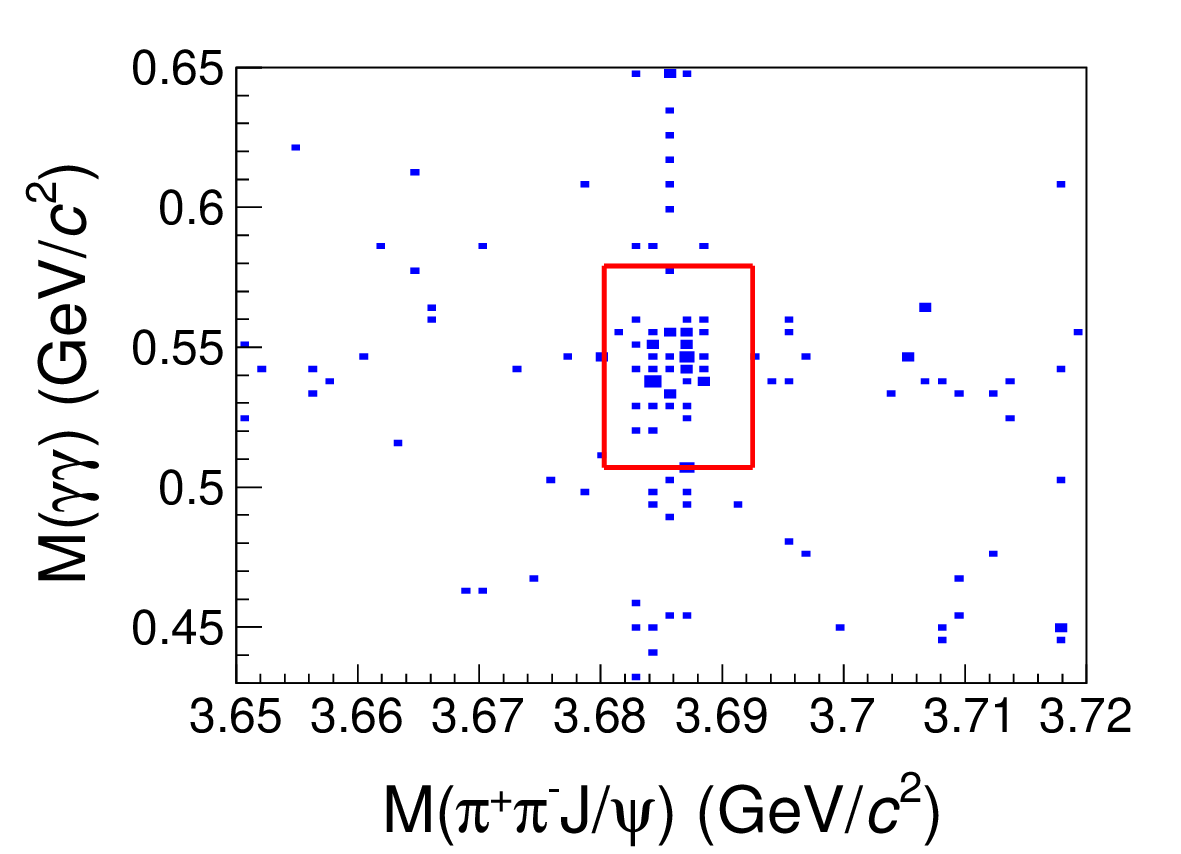}
\put(-115,92){\textbf{(a)}}
 }
\subfigure{
 \label{fig:2d-mass-eta-pipijpsi-f}
\includegraphics[width=0.25\paperwidth]
 {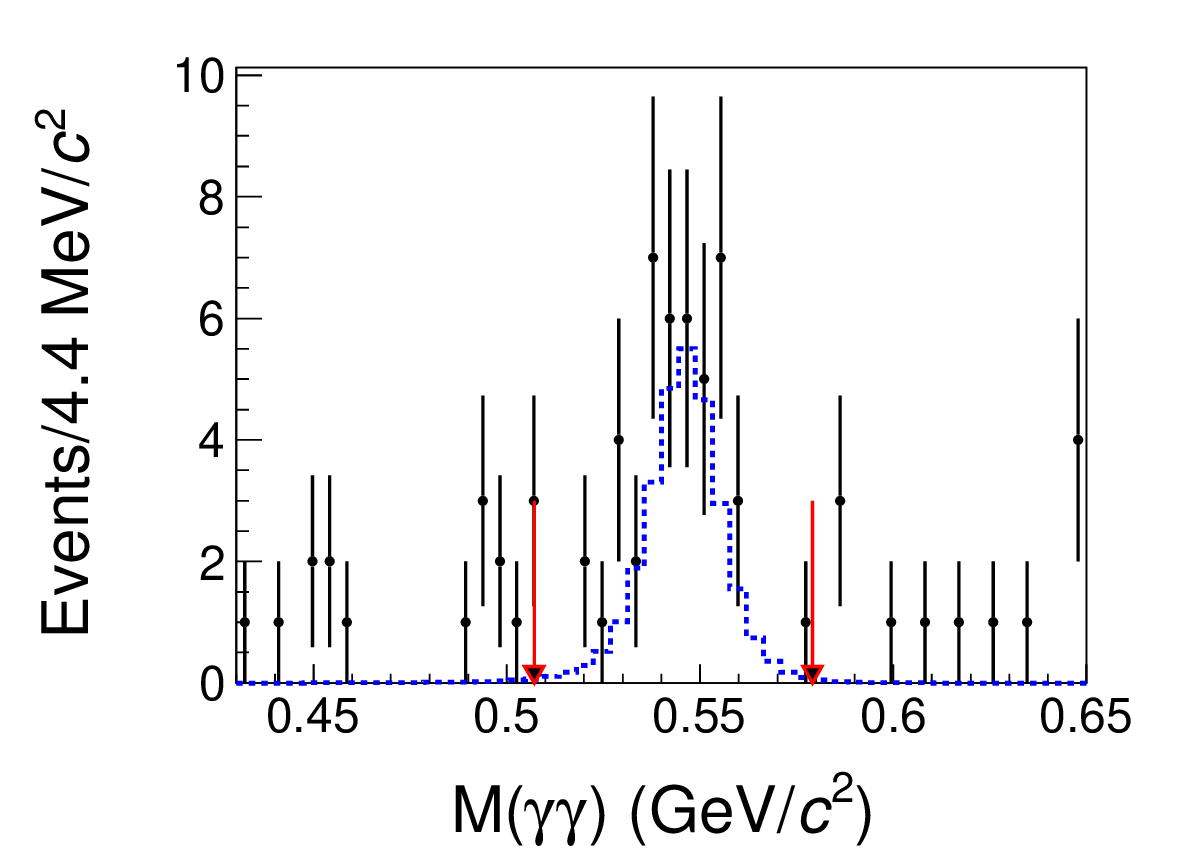}
\put(-115,92){\textbf{(b)}}
 }
 \subfigure{
\label{fig:2d-mass-eta-pipijpsi-g}
\includegraphics[width=0.25\paperwidth]
 {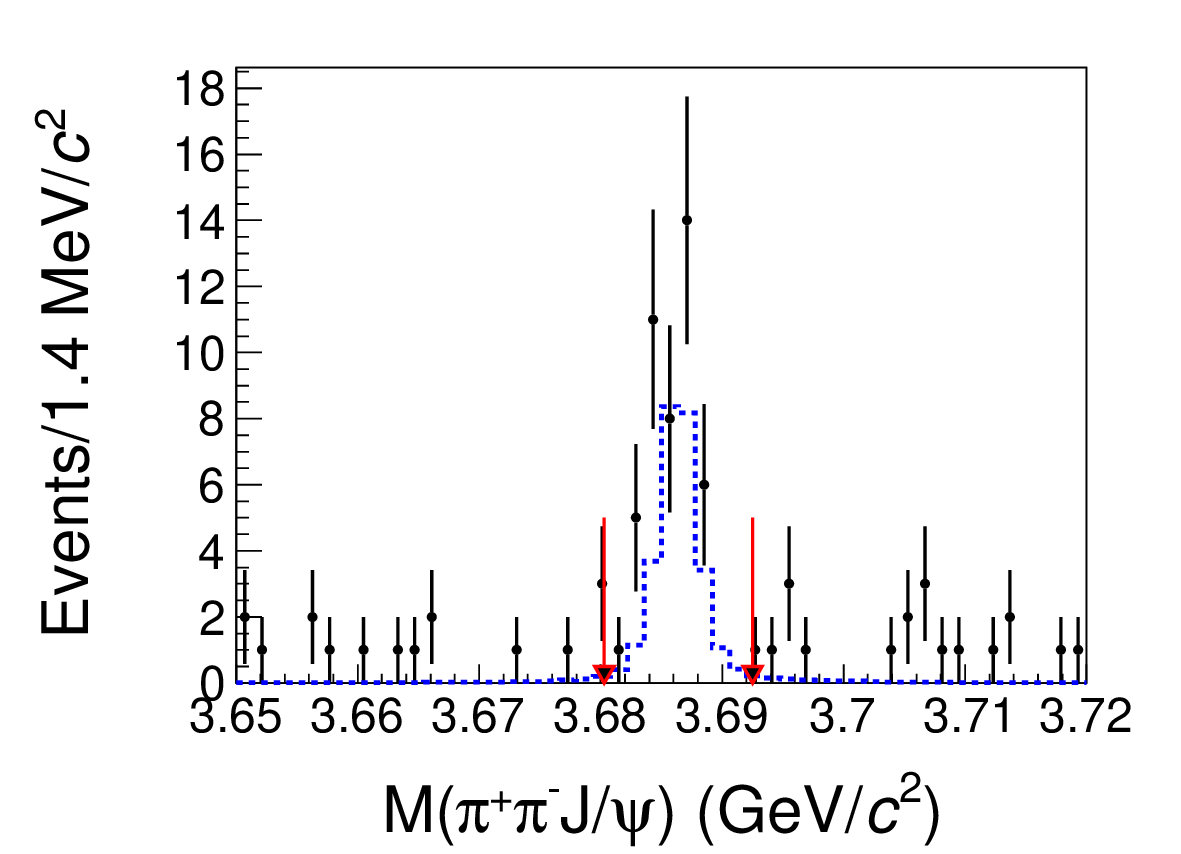}
\put(-115,92){\textbf{(c)}}
 }\\
\caption{\label{fig:2d-mass-eta-pipijpsi}
 (a) Distribution of $M(\gamma\gamma)$ versus $M(\pip\pim \jpsi)$,   (b) the projection along  $M(\gamma\gamma)$ in the $\psip$ mass window, and (c) the projection along   $M(\pip\pim \jpsi)$ in the $\eta$ mass window for the full data set. The red rectangle and arrows represent the mass windows of $\eta$ and $\psip$ selections, and the blue dashed histograms in panels (b) and (c) represent the $\ee\to\eta\psip$ signal MC simulated shapes  of  $M(\gamma\gamma)$  and    $M(\pip\pim \jpsi)$ distributions, respectively. }
\end{figure*}

After applying the aforementioned selection criteria, the invariant mass distributions of $M(\GG)$ versus $M(\ppjpsi)$ for the full data set, for $\ee\to\eta\psip$ signal MC samples,  and the corresponding  projection plots are shown in   Fig.~\ref{fig:2d-mass-eta-pipijpsi}. 
  Here $M(\ppjpsi)=M(\pp\ell^{+}\ell^{-})-M(\ell^{+}\ell^{-})+M(\jpsi)_{\rm PDG}$ is used to eliminate  the detection resolution from  $M(\ell^{+}\ell^{-})$,  and $M(\pp\ell^{+}\ell^{-})$ and $M(\jpsi)_{\rm PDG}$ are the invariant mass of $\pp\ell^{+}\ell^{-}$ and the nominal mass of the $\jpsi$, respectively. 
  The signal regions for the $\eta$ and $\psip$ states are set to be $507.1<M(\gamma\gamma)<579.1$~MeV/$c^2$ and  $3680.3<M(\ppjpsi)<3692.5$~MeV/$c^2$, respectively, quoted from Ref.~\cite{etapsip} and within three times the detector deviations to the nominal masses of $\eta$ and $\psip$ states~\cite{pdg}.  Significant clusters from $\ee\to\eta\psip$ signals can be seen in Fig.~\ref{fig:2d-mass-eta-pipijpsi}.

\subsection{Background analysis}
\label{sec:background1}
\begin{figure*}[tbp]
\renewcommand\figurename{\rm FIG}
\centering 
\subfigure{
\label{fig:2d-mass-eta-pipijpsi-a}
\includegraphics[width=0.25\paperwidth]
 {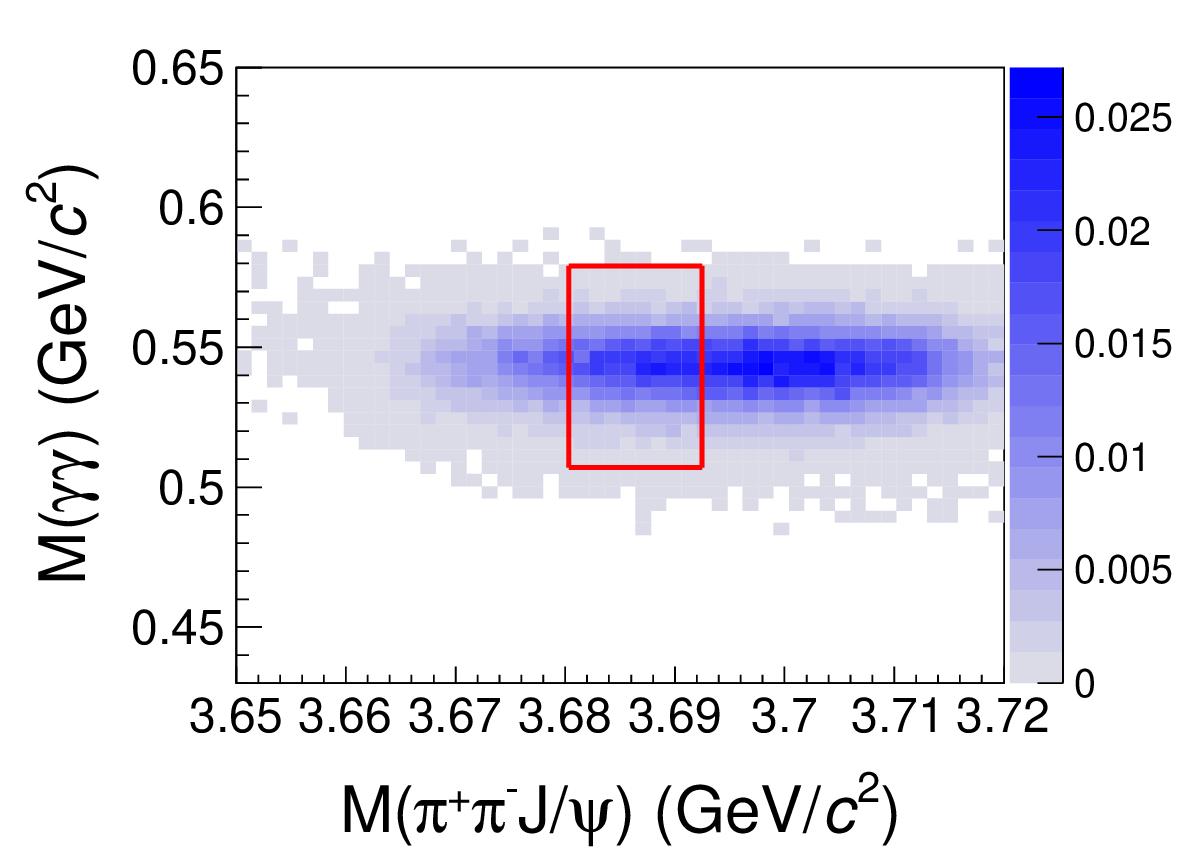}
\put(-115,92){\textbf{(a)}}
 }
\subfigure{
 \label{fig:2d-mass-eta-pipijpsi-f}
\includegraphics[width=0.25\paperwidth]
 {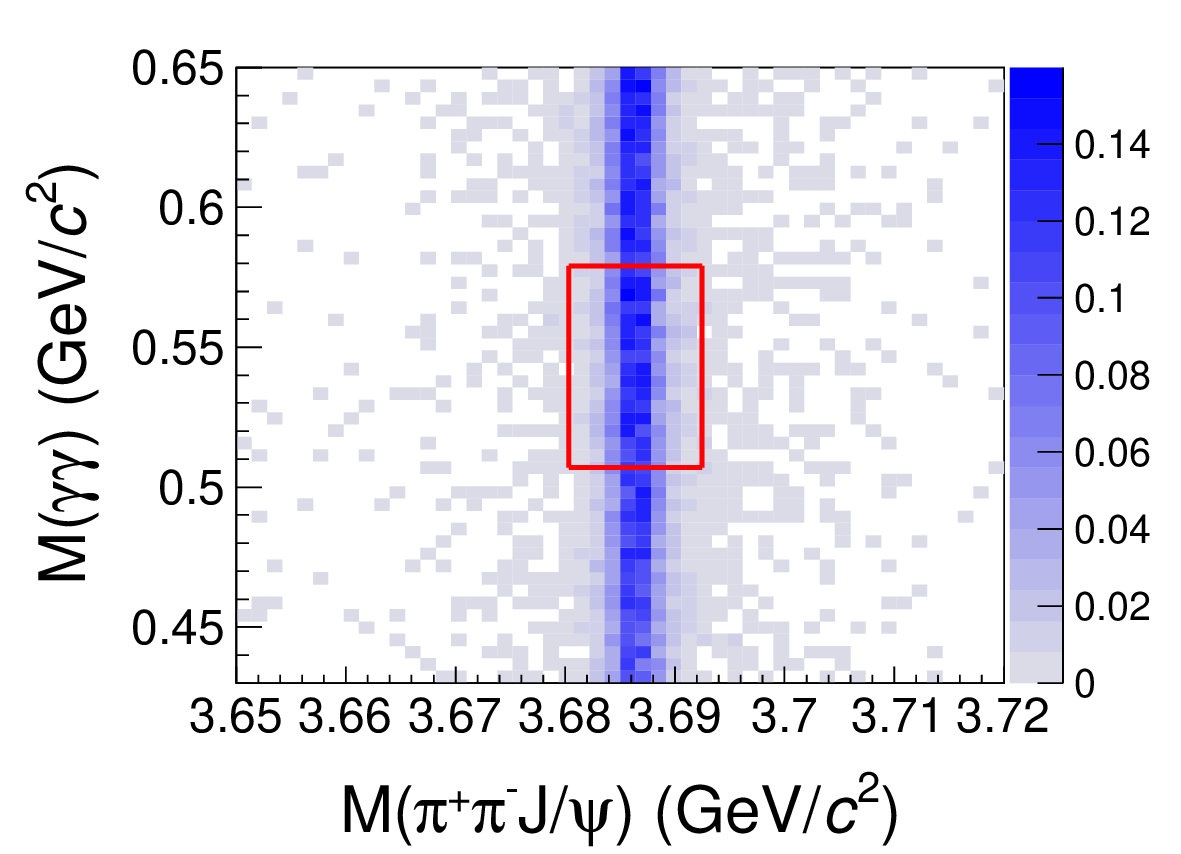}
\put(-115,92){\textbf{(b)}}
 }
 \subfigure{
\label{fig:2d-mass-eta-pipijpsi-g}
\includegraphics[width=0.25\paperwidth]
 {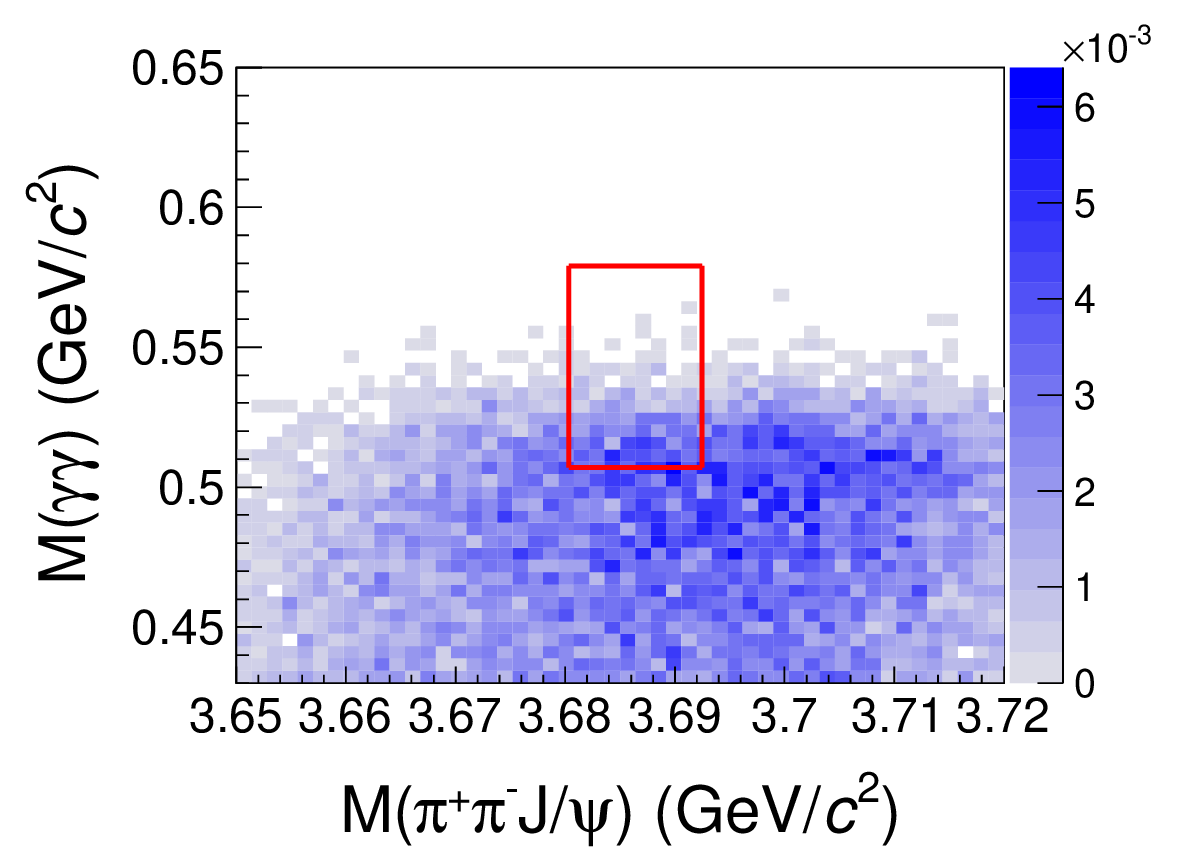}
\put(-115,92){\textbf{(c)}}
 }\\
 \caption{\label{fig:2d-mass-eta-pipijpsi-background}
Distributions of $M(\gamma\gamma)$ versus $M(\pip\pim \jpsi)$ for the exclusive background MC samples of (a) type I: $\ee\to\pppsip$, $\psip\to\jpsi\eta$ at $\sqrt{s}=4.288$ GeV,  (b) type II: $\ee\to\GG\psip$ at $\sqrt{s}=4.682$ GeV, and (c) type III: $\ee\to\pppsip$, $\psip\to\gamma\chi_{c1}$ at $\sqrt{s}=4.288$ GeV. The MC samples are  normalized to the numbers of the observed events in data at the corresponding energy points. The red rectangles  represent the mass windows of $\eta$ and $\psip$ selections.}
\end{figure*}

The same selection criteria are applied to the inclusive
MC sample generated at $\sqrt{s}=4.682$ GeV to investigate possible background contributions.   The corresponding processes are listed in Table~\ref{table:The possible backgrounds}. 
 These backgrounds are divided into three categories: non-$\psi(2S)$ events (type I), non-$\eta$ events (type II), and both non-$\eta$ and non-$\psi(2S)$ events (type III). Figure~\ref{fig:2d-mass-eta-pipijpsi-background} shows the $M(\gamma\gamma)$ versus $M(\pip\pim \jpsi)$ distributions for the three different background categories.

\begin{table*}[htbp]
\renewcommand\tablename{\rm TABLE}
\caption{List of the possible background processes in the $\ee\to\eta\psi(2S)$ signal region~(where $\jpsi\rightarrow\LL$).}\label{table:The possible backgrounds}
\begin{center}
\begin{tabular}{clll}
\hline\hline Type&&Decay mode (Branching fraction)& \\\hline
I&$\EE \to \pip\pim\psi(2S)$,       &$\psi(2S)\to \jpsi \eta$ (3.37\%),             & $\eta \to \gamma \gamma$ (39.41\%)\\ \hline
\multirow{2}{*}{II}
&$\EE \to \pi^{0}\pi^{0}\psi(2S)$, &$\psi(2S)  \to \ppjpsi$  (34.68\%)              &  \\ 
&$\EE \to \gamma\gamma\psi(2S)$, &$\psi(2S)  \to \ppjpsi$  (34.68\%)                 &  \\ \hline
\multirow{4}{*}{III}
&$\EE \to \pip\pim\psi(2S)$,       &$\psi(2S)\to \gamma \chi_{cJ}(J=0,1,2)$ (9.79\%, 9.75\%,  9.52\%),   & $\chi_{cJ} \to \gamma \jpsi$ (1.4\%, 34.3\%, 19.0\%) \\
&$\EE \to \omega  \chi_{cJ}(J=0,1,2)$, &$\omega \to\pip\pim\pi^{0}$  (89.3\%),     & $\chi_{cJ} \to \gamma \jpsi$ (1.4\%, 34.3\%, 19.0\%) \\  
&$\EE \to \phi \chi_{cJ}(J=0,1,2)$,      &$\phi \to \pip \pim \pi^{0}$ (15.24\%),    & $\chi_{cJ} \to \gamma \jpsi$ (1.4\%, 34.3\%, 19.0\%)   \\
&$\EE \to \gamma X(3872)$,        &$X(3872) \to \omega \jpsi$ (2.3\%),      & $\omega \to \pip\pim\pi^{0}$ (89.3\%) \\ \hline\hline
\end{tabular}
\end{center}
\end{table*}

 The dominant background contribution comes from the process $\EE\to\gamma\gamma\psip$ with $\psip\to\ppjpsi$ and $\jpsi\to\LL$, and its yield is measured directly in this analysis.   
After applying all the above selection criteria  
  except from the  $\eta$ mass window,  the $M(\GG)$ distribution in the process $\ee\to\GG\psip$ is shown in Fig.~\ref{fig:2d-mass-eta-pipijpsi-background}(b) as an example. 
   We can measure  its yield  by requiring that the mass  of $\gamma\gamma$ is greater than 300~MeV/$c^2$
and not within $[507.1, 579.1]$~MeV/$c^{2}$. Here, these  mass range requirements are applied to remove the contributions from the  $\ee\to\piz\psip$,  $\ee\to\gamma_{\rm ISR}\psip$, and  $\ee\to\eta\psip$  processes. The contribution from $\ee\to\piz\piz\psip$ is negligible compared to  $\ee\to\GG\psip$. 

The number of $\GG\psip$ events in the $\psip$ signal region but not in the $\eta$  or $\piz$ signal region  [$N_{\gamma\gamma\psip}^0$] is obtained by  substracting   the number of observed events in the $\psip$ sideband regions [$N_{\rm side}^{\rm o}$] from the number of observed events in $\psip$ signal region [$N_{\rm sig}^{\rm o}$], i.e.\ $N_{\gamma\gamma\psip}^0=N_{\rm sig}^{\rm o}-N_{\rm side}^{\rm o}$.
 The number of $\ee\to \gamma\gamma\psip$ events in the $\eta$ and $\psip$ signal regions [$N_{\gamma\gamma\psip}^1$] is  obtained from the $N_{\gamma \gamma \psi(2S)}^{0}$ as follows:
\begin{equation}\label{equ:bk2 number}
N_{\gamma\gamma\psip}^1 = F\cdot N_{\gamma \gamma \psi(2S)}^{0},
\end{equation}

We define a factor $F$ from branching fractions and selection efficiencies, as: 
\begin{equation}\label{equ:f factor}
F = \frac{\epsilon^{1}_{e}\mathcal{B}_{e}+\epsilon_{\mu}^{1}\mathcal{B}_\mu}{\epsilon^{0}_{e}\mathcal{B}_{e}+\epsilon_{\mu}^{0}\mathcal{B}_\mu},
\end{equation}
 where $\epsilon^{1}_{e}$ and $\epsilon_{\mu}^{1}$ are the detection efficiencies of $\ee\to\gamma\gamma\psip$ background events for the 
$\jpsi\to\EE$ and $\jpsi\to\MM$ decay channels in the $\eta$ and $\psip$ signal regions, respectively; $\epsilon^{0}_{e}$ and $\epsilon_{\mu}^{0}$ are the detection efficiencies of $\ee\to\gamma\gamma\psip$ background events for the 
$\jpsi\to\EE$ and $\jpsi\to\MM$ decay channels in the $\psip$ signal region but outside the $\eta$ and $\piz$ signal regions, respectively;  $\mathcal{B}_e$ and $\mathcal{B}_\mu$ are the branching fractions of decays  $\jpsi\to\EE$ and $\jpsi\to\MM$, respectively~\cite{pdg}. 
The values related to the $\ee\to\GG\psip$  backgrounds are listed in Table~\ref{table:results for gamma gamma psip}.  The uncertainty on $F$ is neglected since it is very small.

\begin{table}[htbp]
\renewcommand\tablename{\rm TABLE}
\caption{The numbers of $\ee\to \gamma\gamma\psi(2S)$ events outside [$N_{\gamma \gamma \psi(2S)}^{0}$] and inside [$N_{\gamma \gamma \psi(2S)}^{1}$] the $\eta$  signal region,  the $F$ factor,  and the numbers of observed events in  the signal [$N_{\rm sig}^{\rm o}$] and sideband [$N_{\rm side}^{\rm o}$] regions of the $\psi(2S)$ mass window at each c.m.\ energy. The statistical uncertainty of $N_{\gamma \gamma \psi(2S)}^{0}$ is calculated by the Feldman-cousins method~\cite{feldman-cousins-method}.}\label{table:results for gamma gamma psip}
\begin{center}
\begin{tabular}{cccccc}
\hline\hline
$\sqrt s$~(GeV) &F&$N_{\rm sig}^{\rm o}$&$N_{\rm side}^{\rm o}$&$N_{\rm \gamma\gamma\psi(2S)}^0$&$N_{\gamma\gamma\psi(2S)}^1$\\\hline
    4.288  & 0.246 &   6  &  6  & $ 0.00_{-0.00}^{+3.28}$ &   $0.00_{-0.00}^{+0.81}$\\
    4.312  & 0.265 &   4  &  0  & $ 4.00_{-1.66}^{+2.78}$ &   $1.06_{-0.44}^{+0.74}$\\
    4.337  & 0.265 &   7  &  0  & $ 7.00_{-2.74}^{+3.31}$ &   $1.86_{-0.73}^{+0.88}$\\
    4.377  & 0.243 &   2  &  1  & $ 1.00_{-0.86}^{+2.26}$ &   $0.24_{-0.21}^{+0.55}$\\
    4.396  & 0.237 &  10  &  0  & $10.00_{-3.22}^{+3.81}$ &   $2.37_{-0.76}^{+0.90}$\\
    4.436  & 0.209 &   8  &  0  & $ 8.00_{-2.70}^{+3.32}$ &   $1.67_{-0.56}^{+0.69}$\\
    4.612  & 0.141 &   1  &  0  & $ 1.00_{-0.63}^{+1.76}$ &   $0.14_{-0.09}^{+0.25}$\\
    4.628  & 0.134 &   4  &  1  & $ 3.00_{-1.66}^{+2.78}$ &   $0.40_{-0.22}^{+0.37}$\\
    4.641  & 0.136 &   2  &  0  & $ 2.00_{-1.26}^{+2.26}$ &   $0.27_{-0.17}^{+0.31}$\\
    4.661  & 0.125 &   4  &  0  & $ 4.00_{-1.66}^{+2.78}$ &   $0.50_{-0.21}^{+0.35}$\\
    4.682  & 0.126 &  19  &  0  & $19.00_{-4.18}^{+4.83}$ &   $2.39_{-0.53}^{+0.61}$\\
    4.699  & 0.120 &   2  &  0  & $ 2.00_{-1.26}^{+2.26}$ &   $0.24_{-0.15}^{+0.27}$\\
    4.740  & 0.110 &   0  &  0  & $ 0.00_{-0.00}^{+1.29}$ &   $0.00_{-0.00}^{+0.14}$\\
    4.750  & 0.106 &   5  &  0  & $ 5.00_{-2.24}^{+2.81}$ &   $0.53_{-0.24}^{+0.30}$\\
    4.781  & 0.100 &   7  &  0  & $ 7.00_{-2.74}^{+3.31}$ &   $0.70_{-0.27}^{+0.33}$\\
    4.843  & 0.087 &   5  &  1  & $ 4.00_{-2.24}^{+2.81}$ &   $0.35_{-0.20}^{+0.25}$\\
    4.918  & 0.077 &   4  &  0  & $ 4.00_{-1.66}^{+2.78}$ &   $0.31_{-0.13}^{+0.21}$\\
    4.951  & 0.075 &   0  &  0  & $ 0.00_{-0.00}^{+1.29}$ &   $0.00_{-0.00}^{+0.10}$\\
  \hline\hline
\end{tabular}
\end{center}
\end{table}

 The yields for each of the other background processes in the signal region ($N_{{\rm bkg},i}$) are calculated using:
\begin{equation}\label{equ:bk number}
\begin{split}
\begin{aligned}
 N_{{\rm bkg},i}=\mathcal{L}_{\rm int}(1+\delta)_i |1-\Pi|^{-2}
\epsilon_i\mathcal{B}_i\sigma^{\rm B}_{{\rm bkg},i},
  \end{aligned}
\end{split}
\end{equation}
where $i$ represents each background channel; $\mathcal{L}_{\rm int}$ is the integrated luminosity; $|1-\Pi|^{-2}$ is the vacuum polarization factor~\cite{ vacuum-polarization-factor}; $\epsilon_i$ and $\mathcal{B}_i$ are the selection efficiency and the product branching fraction of the intermediate states taken from
the PDG~\cite{pdg} for the $i^{\rm th}$ background mode, respectively; and $\sigma^{\rm B}_{{\rm bkg},i}$ is the measured Born cross section of the $i^{\rm th}$ background mode.
The production cross sections for these background processes are  taken from Refs.~\cite{cross-section-pppsip,cross-section-p0p0psip,cross-section-omegachic012, cross-section-omegachic012-2, bes3-Y4230-omegachic0, cross-section-gammax3872,cross-section-phichic12}. 
Assuming an input line shape from Refs.~\cite{cross-section-pppsip,cross-section-p0p0psip,cross-section-omegachic012, cross-section-omegachic012-2, bes3-Y4230-omegachic0,cross-section-gammax3872,cross-section-phichic12}, the ISR correction factor $(1+\delta)_i$ is obtained from a quantum electrodynamics calculation~\cite{isr-calculate2} using the {\sc kkmc} generator~\cite{ref-kkmc, ref-kkmc-2}.

The total number of background events ($n^{\rm b}$)  in the signal region of $\ee\to\eta\psip$ is obtained by
\begin{equation}\label{equ:bk number2}
\begin{split}
\begin{aligned}
 n^{\rm b}&=\sum_i N_{{\rm bkg},i}+N_{{\gamma\gamma\psip}}^1  .\\
   \end{aligned}
\end{split}
\end{equation}
Finally, the total numbers of background events in the signal region at different energy points, together with the numbers of background events from different
decay processes, are listed in Table~\ref{table:The expected events from different background MC samples}.   The uncertainties for the numbers of the  $\ee\to\gamma\gamma\psip$ events are statistical only, while for  the other backgrounds events, they are the statistical and systematic uncertainties added in quadrature.

\begin{table*}[htbp]
\renewcommand\tablename{\rm TABLE}
\caption{The total numbers of background events  in the signal region ($n^{\rm b}$) at different energy points for $\ee\to\eta\psi(2S)$ signal process, together with the expected numbers of background events from different processes.  Ellipses mean that the results are  0 or close to 0.  Here the $\ee\to\gamma\gamma\psip$ uncertainties are statistical only, while for  the other backgrounds are the  sum of the statistical and the systematic uncertainties in quadrature.}
\label{table:The expected events from different background MC samples} 
\begin{center}
\begin{tabular}{ccccccc}
\hline\hline
 $\sqrt s$ (GeV)& 4.288& 4.312& 4.337& 4.377& 4.396& 4.436\\\hline
        $\pip\pim\psi(2S),\psi(2S)\to\jpsi\eta$      &$ 1.07 \pm 0.09 $&$ 0.23 \pm 0.01 $&$ 0.02$&$\cdots$&$\cdots$&$\cdots$\\
        $\pip\pim\psi(2S),\psi(2S)\to\gamma\chi_{c0}$&$ 0.02$&$ 0.01$&$\cdots$&$\cdots$&$\cdots$&$\cdots$\\
        $\pip\pim\psi(2S),\psi(2S)\to\gamma\chi_{c1}$&$ 0.15 \pm 0.01 $&$ 0.07$&$ 0.01$&$\cdots$&$\cdots$&$\cdots$\\
        $\pip\pim\psi(2S),\psi(2S)\to\gamma\chi_{c2}$&$\cdots$&$\cdots$&$\cdots$&$\cdots$&$\cdots$&$\cdots$\\
        $\pi^{0}\pi^{0}\psi(2S)$                     &$ 0.05 \pm 0.01 $&$ 0.07 \pm 0.01 $&$ 0.11 \pm 0.01 $&$ 0.19 \pm 0.02 $&$ 0.22 \pm 0.03 $&$ 0.21 \pm 0.03$\\
        $\omega \chi_{c0}$                           &$\cdots$&$\cdots$&$\cdots$&$\cdots$&$\cdots$&$\cdots$\\
        $\omega \chi_{c1}$                           &$\cdots$&$ 0.01 \pm 0.01 $&$ 0.01 \pm 0.01 $&$ 0.01 \pm 0.01 $&$ 0.01$&$ 0.01 \pm 0.01$\\
        $\omega \chi_{c2}$                           &$\cdots$&$\cdots$&$\cdots$&$ 0.05 \pm 0.02 $&$ 0.06 \pm 0.01 $&$ 0.06 \pm 0.01$\\
        $\gamma X(3872)$           &$\cdots$&$\cdots$&$\cdots$&$\cdots$&$\cdots$&$\cdots$\\

  $\gamma\gamma\psi(2S)$                       &$0.00   ^{+0.81  }_{-0.00  }$&$1.06   ^{+0.74  }_{-0.44  }$&$1.86   ^{+0.88  }_{-0.73  }$&$0.24   ^{+0.55}_{-0.21  }$&$2.37   ^{+0.90  }_{-0.76  }$&$1.67   ^{+0.69  }_{-0.56}$\\\hline
  $n^{\rm b}$                                  &$ 1.29 \pm 0.81 $&$ 1.45 \pm 0.74 $&$ 2.00 \pm 0.88 $&$ 0.50 \pm 0.55 $&$ 2.65 \pm 0.90 $&$ 1.95 \pm 0.69$\\\hline 
 $\sqrt s$ (GeV)& 4.612& 4.628& 4.641& 4.661& 4.682& 4.699\\\hline
      $\phi \chi_{cJ}$           &$\cdots$&$\cdots$&$\cdots$&$\cdots$&$\cdots$&$\cdots$\\
$\pi^{0}\pi^{0}\psi(2S)$   & $ 0.01 \pm 0.00 $&$ 0.07 \pm 0.01 $&$ 0.08 \pm 0.01 $&$ 0.10 \pm 0.02 $&$ 0.31 \pm 0.04 $&$ 0.07 \pm 0.01$\\
      $\omega \chi_{cJ}$           &$\cdots$&$\cdots$&$\cdots$&$\cdots$&$\cdots$&$\cdots$\\
      $\gamma X(3872)$           &$\cdots$&$\cdots$&$\cdots$&$\cdots$&$\cdots$&$\cdots$\\
  $\gamma\gamma\psi(2S)$                       &$0.14   ^{+0.25  }_{-0.09  }$&$0.40   ^{+0.37  }_{-0.22  }$&$0.27   ^{+0.31  }_{-0.17  }$&$0.50   ^{+0.35}_{-0.21  }$&$2.39   ^{+0.61  }_{-0.53  }$&$0.24   ^{+0.27  }_{-0.15}$\\\hline
 $n^{\rm b}$                & $ 0.15 \pm 0.25 $&$ 0.48 \pm 0.37 $&$ 0.36 \pm 0.31 $&$ 0.61 \pm 0.35 $&$ 2.70 \pm 0.61 $&$ 0.31 \pm 0.27$\\\hline 
$\sqrt s$ (GeV)& 4.740& 4.750& 4.781& 4.843& 4.918& 4.951\\\hline
      $\phi \chi_{cJ}$           &$\cdots$&$\cdots$&$\cdots$&$\cdots$&$\cdots$&$\cdots$\\$\pi^{0}\pi^{0}\psi(2S)$&$ 0.02 \pm 0.01 $&$ 0.03 \pm 0.03 $&$ 0.04 \pm 0.03 $&$ 0.02 \pm 0.03 $&$ 0.01 \pm 0.01 $&$\cdots$\\
      $\omega \chi_{cJ}$           &$\cdots$&$\cdots$&$\cdots$&$\cdots$&$\cdots$&$\cdots$\\
      $\gamma X(3872)$           &$\cdots$&$\cdots$&$\cdots$&$\cdots$&$\cdots$&$\cdots$\\
 $\gamma\gamma\psi(2S)$                       &$0.00   ^{+0.14  }_{-0.00  }$&$0.53   ^{+0.30  }_{-0.24  }$&$0.70   ^{+0.33  }_{-0.27  }$&$0.35   ^{+0.24}_{-0.20  }$&$0.31   ^{+0.21  }_{-0.13  }$&$0.00   ^{+0.10  }_{-0.00}$\\\hline

 $n^{\rm b}$                &$ 0.02 \pm 0.14 $&$ 0.56 \pm 0.30 $&$ 0.74 \pm 0.33 $&$ 0.37 \pm 0.25 $&$ 0.31 \pm 0.21 $&$ 0.00 \pm 0.10$\\\hline\hline

\end{tabular}
\end{center}
\end{table*}

\subsection{Cross section measurement}
\label{sec:cross1}

Since there are only a few events in the signal region of $\ee\to\eta\psip$, 
 the number of observed events ($n^{\rm obs}$), which is the sum of the number of expected background
($n^{\rm  b}$) and signal ($x$) events, follows a Poisson distribution,
\begin{equation}\label{equ:Poisson}
  P(n^{\rm obs};x, n^{\rm b})=\frac{(x+ n^{\rm b})^{n^{\rm obs}}}{n^{\rm obs}!}e^{-(x+ n^{\rm b})}.
\end{equation}
In this analysis, the number of signal events $\ee\to\eta\psip$ is obtained using the  likelihood ratio ordering (also known as Feldman–Cousins, F-C)  method~\cite{feldman-cousins-method}, i.e.  the value of $x$    corresponding to    the maximum $P(n^{\rm obs};  x, n^{\rm b})$.  Thus, $n^{\rm sig} = {\rm max}(0,  n^{\rm obs}- n^{\rm b})$ is the best estimation of the number of signal events in the physically-allowed region.

The statistical uncertainty of the number of signal events at a 68.27\% confidence level (C.L.) is estimated with the F-C method~\cite{feldman-cousins-method}. Since no significant $\eta\psip$ signals  are observed at some of the energy points, the confidence intervals for the number of signal events with the lower and upper limits at a 90\% C.L.\  are obtained with  a toolkit which supports the F-C method  and takes into account the systematic uncertainty, the Poissonian limit estimator (POLE)  program~\cite{pole-method}.

The Born cross section of $\EE\to\eta\psip$ is calculated by
\begin{equation}
\begin{split}
\begin{aligned}
 \sigma^{\rm B}[\ee\to\eta\psip]= \\
 \frac{n^{\rm sig}}
 {\mathcal{L_{\rm int}}(1+\delta)|1-\Pi|^{-2}
   \mathcal{B}_1
   \mathcal{B}_2
  (\epsilon_e\mathcal{B}_e+\epsilon_\mu\mathcal{B}_\mu)
   },
  \end{aligned}
   \end{split}
\end{equation} 
where $\mathcal{B}_1$ and $\mathcal{B}_2$ are the branching fractions of $\psip\to\ppjpsi$ and $\eta\to\gamma\gamma$~\cite{pdg}, respectively; $\epsilon_e$ and $\epsilon_\mu$ are the detection efficiencies for $\EE$ and $\MM$ modes, respectively;     $(1+\delta)$ is the radiative correction factor obtained from  the quantum electrodynamics calculation~\cite{isr-calculate2} using the {\sc kkmc} generator~\cite{ref-kkmc, ref-kkmc-2}, assuming that the cross section follows  an input line shape of the $Y(4260)$~\cite{pdg} at the energy points below 4.600 GeV, and an input line shape in form of a   power function $1/s$ at the energy points above 4.600 GeV. 
The Born cross sections, the corresponding  confidence intervals with the lower and upper limits at  90\% C.L., and the numbers used in the calculation are listed in Table~\ref{table:Preliminary results}.

\begin{table*}[htbp]
\renewcommand\tablename{\rm TABLE}
 \caption{ 
The cross sections $\sigma^{\rm B}$ and the confidence intervals with the lower and upper limits on $\sigma^{\rm B}$  with the POLE method ($\sigma^{\rm B}_{\rm POLE}$) 
for $\EE\to \eta\psip$ at different energy points, together with integrated luminosities $\mathcal{L}_{\rm int}$, the numbers of observed events $n^{\rm obs}$,  the background events $n^{\rm b}$,  the signal events $n^{\rm sig}$,  the confidence intervals with the lower and upper limits 
 for the numbers of signal events $n^{\rm sig}_{\rm POLE}$, the products of detection efficiencies and branching fractions $\Sigma=\mathcal{B}_1\mathcal{B}_2 (\epsilon_e\mathcal{B}_e+\epsilon_\mu\mathcal{B}_\mu)$,  and the
products of  ISR correction factor  and  vacuum polarization factor $(1+\delta)|1-\Pi|^{-2}$. 
The uncertainties of $n^{\rm sig}$ and $\sigma^{\rm B}$ are statistical only. The lower and upper limits  are given at 90\%
C.L.\ including the systematic uncertainties. 
}\label{table:Preliminary results}
 \begin{center}
\begin{tabular}{cccccccccc}
\hline\hline
$\sqrt s$~(GeV) &$ \mathcal{L}_{\rm int}$~(pb$^{-1}$)&$n^{\rm obs}$&$n^{\rm b}$ &$n^{\rm sig}$&$n^{\rm sig}_{\rm POLE}$
& $\Sigma~(10^{-2})$ &$(1+\delta)|1-\Pi|^{-2}$ & $\sigma^{B}$~(pb)  & $\sigma^{\rm B}_{\rm POLE}$~(pb)\\\hline
 4.288&    491.5&  6&  1.29&              $4.7_{-2.2}^{+3.3}$&           (1.4,~0.1)&    0.363&  0.90&                   $2.9_{-1.3}^{+2.0}$&           (0.9,~6.3)\\
    4.312&    492.1&  4&  1.45&              $2.5_{-1.6}^{+2.8}$&           (0.4,~7.5)&    0.333&  1.01&                   $1.5_{-1.0}^{+1.7}$&           (0.2,~4.5)\\
    4.337&    501.1&  4&  2.00&              $2.0_{-1.6}^{+2.8}$&           (0.0,~7.2)&    0.301&  1.13&                   $1.2_{-0.9}^{+1.6}$&           (0.0,~4.2)\\
    4.377&    522.8&  6&  0.50&              $5.5_{-2.2}^{+3.3}$&           (1.8,~12.0)&    0.255&  1.34&                   $3.1_{-1.2}^{+1.8}$&           (1.0,~6.8)\\
    4.396&    505.0&  3&  2.65&              $0.3_{-0.3}^{+2.3}$&           (0.0,~5.2)&    0.236&  1.44&                   $0.2_{-0.2}^{+1.3}$&           (0.0,~3.1)\\
    4.436&    568.1&  9&  1.95&              $7.1_{-2.7}^{+3.8}$&           (2.5,~14.3)&    0.207&  1.65&                   $3.6_{-1.4}^{+2.0}$&           (1.3,~7.4)\\
   4.612&    103.7&  0&  0.15&              $0.0_{-0.0}^{+1.1}$&           (0.0,~2.4)&    0.296&  0.97&                   $0.0_{-0.0}^{+3.9}$&           (0.0,~8.2)\\
    4.628&    521.5&  0&  0.48&              $0.0_{-0.0}^{+0.8}$&           (0.0,~2.4)&    0.293&  0.97&                   $0.0_{-0.0}^{+0.6}$&           (0.0,~1.6)\\
    4.641&    551.7&  2&  0.36&              $1.6_{-1.1}^{+2.3}$&           (0.1,~5.5)&    0.294&  0.98&                   $1.0_{-0.7}^{+1.4}$&           (0.1,~3.4)\\
    4.661&    529.4&  2&  0.61&              $1.4_{-1.0}^{+2.3}$&           (0.0,~5.3)&    0.289&  0.98&                   $0.9_{-0.7}^{+1.5}$&           (0.0,~3.6)\\
    4.682&   1667.4&  3&  2.70&              $0.3_{-0.3}^{+2.3}$&           (0.0,~4.9)&    0.290&  0.98&                   $0.1_{-0.1}^{+0.5}$&           (0.0,~1.0)\\
    4.699&    535.5&  1&  0.31&              $0.7_{-0.6}^{+1.8}$&           (0.0,~4.0)&    0.291&  0.98&                   $0.5_{-0.4}^{+1.1}$&           (0.0,~2.6)\\
    4.740&    163.9&  0&  0.02&              $0.0_{-0.0}^{+1.3}$&           (0.0,~2.5)&    0.297&  0.99&                   $0.0_{-0.0}^{+2.7}$&           (0.0,~5.2)\\
    4.750&    366.5&  3&  0.56&              $2.4_{-1.7}^{+2.3}$&           (0.5,~6.8)&    0.296&  0.99&                   $2.3_{-1.5}^{+2.1}$&           (0.5,~6.3)\\
    4.781&    511.5&  1&  0.74&              $0.3_{-0.3}^{+1.8}$&           (0.0,~3.8)&    0.295&  0.99&                   $0.2_{-0.2}^{+1.2}$&           (0.0,~2.5)\\
    4.843&    525.2&  1&  0.37&              $0.6_{-0.6}^{+1.8}$&           (0.0,~4.0)&    0.294&  1.00&                   $0.4_{-0.4}^{+1.1}$&           (0.0,~2.6)\\
    4.918&    207.8&  0&  0.31&              $0.0_{-0.0}^{+1.0}$&           (0.0,~2.4)&    0.288&  1.01&                   $0.0_{-0.0}^{+1.6}$&           (0.0,~4.0)\\
    4.951&    159.3&  0&  0.00&              $0.0_{-0.0}^{+1.3}$&           (0.0,~2.4)&    0.285&  1.01&                   $0.0_{-0.0}^{+2.8}$&           (0.0,~5.3)\\
   \hline\hline
 \end{tabular}
\end{center} 
\end{table*}

\begin{figure}[htbp]
\renewcommand\figurename{\rm FIG}
\begin{center}
\includegraphics[width=3.0in,height=2in]
 {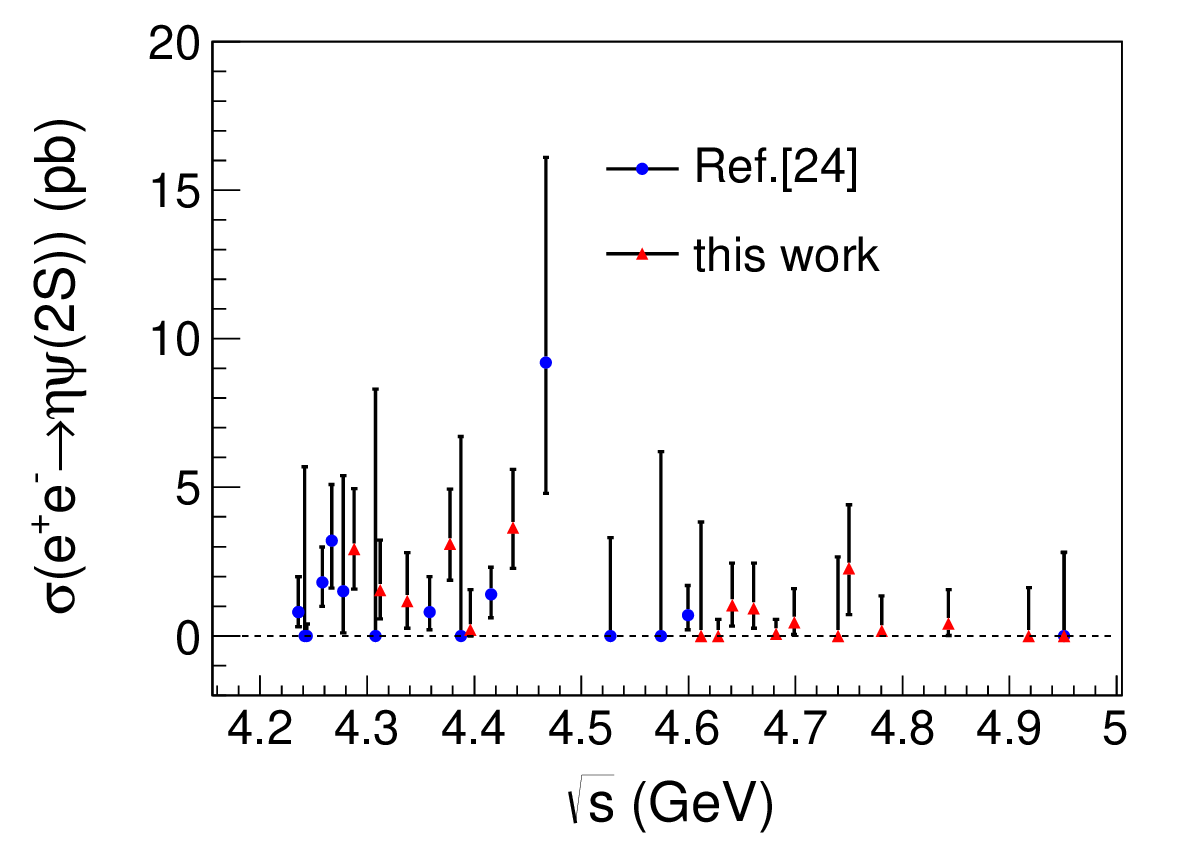} 
\end{center}
\caption{ The measured Born cross sections as a function of the c.m.\ energy.  The blue dots and red triangles are from Ref.~\cite{etapsip} and this measurement, respectively. The error bars represent the statistical errors only.   
 }   
\label{fig:cross section}
\end{figure}

Figure~\ref{fig:cross section} shows the measured Born cross sections for
$\EE\to\eta\psip$  as a function of the c.m.\ energy. 
Due to the limited statistics, it is difficult to draw a clear conclusion whether the vector charmonium-like states exist in the cross section distribution or not.  

\section{\boldmath  Analysis of  $\ee\to\eta\tx$}
\label{sec:de2}

\subsection{Event selection}
\label{sec:eventselection2}
Since we search for the $\tx$ signals in the $\ee\to\eta\tx\to\eta\ppjpsi$ process, with $\eta\to\GG$ and $\jpsi\to\EE$ or $\MM$,
the applied selections are similar to the ones used for the $\ee\to\eta\psip$ process, except for the signal and sideband regions of $\tx$ applied in the $M(\ppjpsi)$ distribution.

\begin{figure*}[tbp]
\renewcommand\figurename{\rm FIG}
\centering 
\subfigure{
\label{fig:2d-mass-eta-pipijpsi-a}
\includegraphics[width=0.25\paperwidth]
 {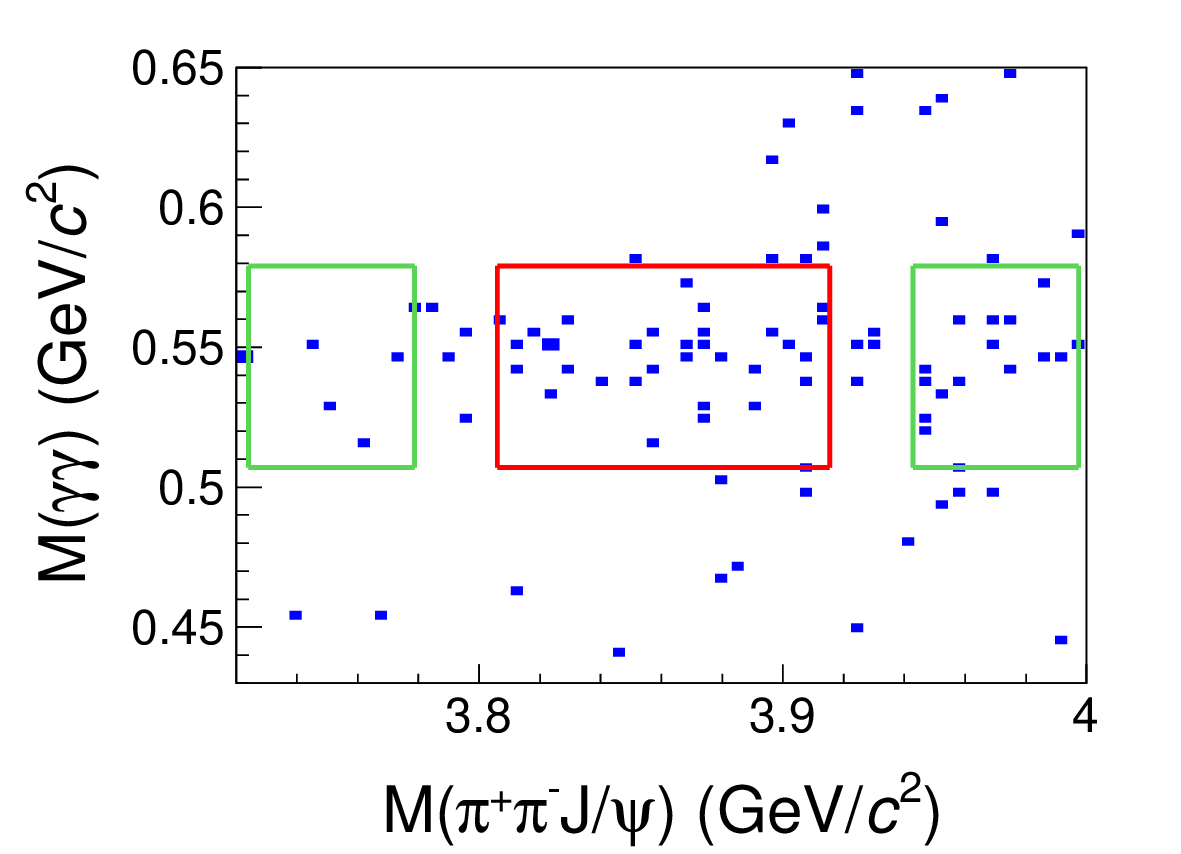}
\put(-115,92){\textbf{(a)}}
 }
\subfigure{
 \label{fig:2d-mass-eta-pipijpsi-f}
\includegraphics[width=0.25\paperwidth]
 {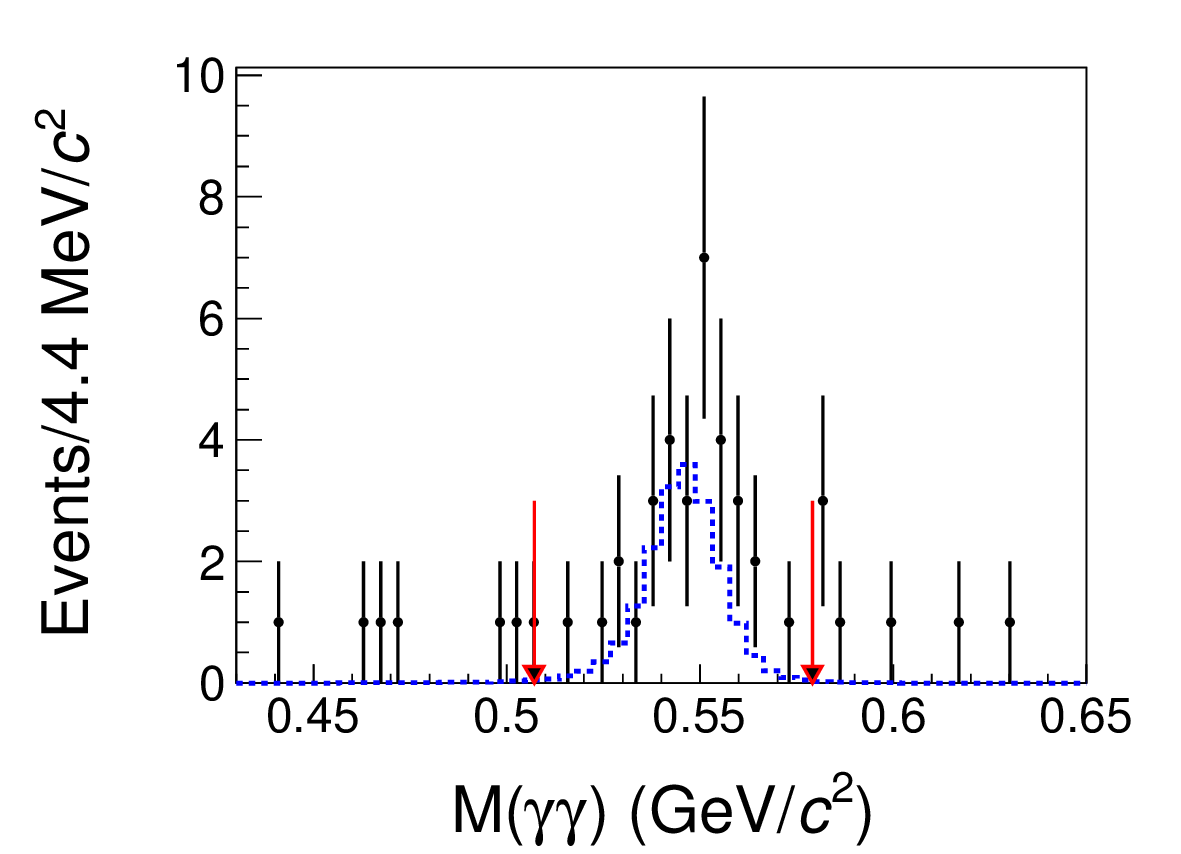 }
\put(-115,92){\textbf{(b)}}
 }
 \subfigure{
\label{fig:2d-mass-eta-pipijpsi-g}
\includegraphics[width=0.25\paperwidth]
 {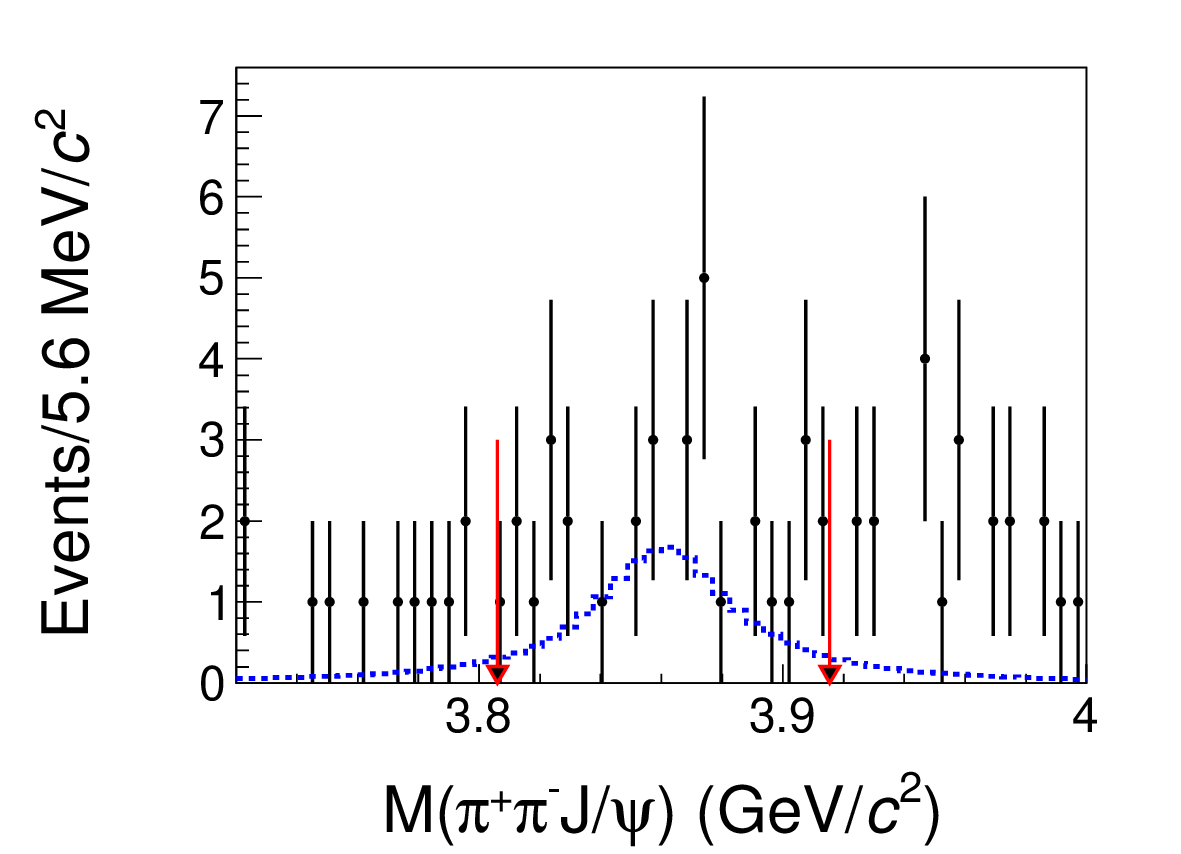 }
\put(-115,92){\textbf{(c)}}
 }\\ 
\caption{\label{fig:scatter plot of etax3872}
 (a) Distribution of $M(\gamma\gamma)$ versus $M(\pip\pim \jpsi)$,   (b) the projection along  $M(\gamma\gamma)$ in the $\tx$ mass window, and (c) the projection along   $M(\pip\pim \jpsi)$ in the $\eta$ mass window for the full data sample. The red rectangle and arrows represent the signal region, the green rectangles represent the  sideband regions, and the blue dashed histograms in panels (b) and (c) represent the $\ee\to\eta\tx$ signal MC simulated shapes  of  $M(\gamma\gamma)$  and    $M(\pip\pim \jpsi)$ distributions for setting I, respectively. }
\end{figure*}

The $\tx$ signal region in the $M(\ppjpsi)$ distribution for the setting I (setting II) is required to be within two (three) times the detector resolution from the COMPASS $\tx$ mass~\cite{compass} (PDG $\tx$ mass~\cite{pdg}),  $3805.9<M(\ppjpsi)<3915.9$ MeV/$c^{2}$ ($3849.5<M(\ppjpsi)<3871.3$ MeV/$c^{2}$).   
The  sideband regions  are required to be within the range from three (four) to five (seven) times the  detector resolution around  the known $\tx$ mass,  $3723.5< M(\ppjpsi)< 3778.4$ MeV/$c^{2}$ and $3943.4<M(\ppjpsi)< 3998.4$ MeV/$c^{2}$  ($3835.1<M(\ppjpsi)< 3845.9$ MeV/$c^{2}$  and $3874.9<M(\ppjpsi)< 3885.7$ MeV/$c^{2}$). Figure~\ref{fig:scatter plot of etax3872} shows the $M(\gamma\gamma)$ and $M(\ppjpsi)$  distributions for the full data sample,   together with the corresponding  projection plots. There is no evidence for $\ee\to\eta\tx$ under either of the $\tx$ width assumptions. 

\subsection{Background analysis}
\label{sec:background2}

\begin{figure}[tbp]
\renewcommand\figurename{\rm FIG}
\begin{center} 
\includegraphics[width=0.3\paperwidth]
 {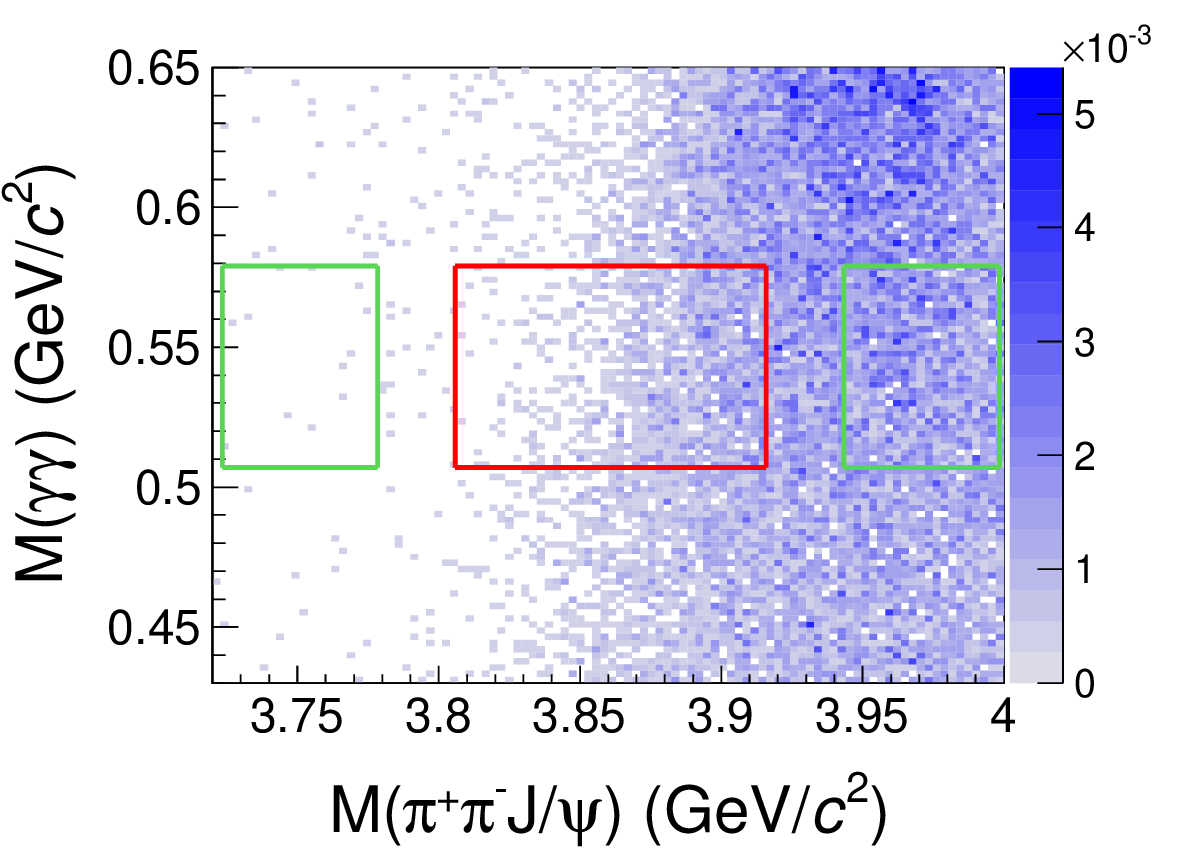}
\end{center}
\caption{Distribution of $M(\gamma\gamma)$ versus $M(\pip\pim \jpsi)$ from $\ee\to\pp\psi(3823)$, $\psi(3823)\to\gamma\chi_{c1}$ background MC sample  at $\sqrt{s}=4.682$ GeV, normalized to the number of the observed events of  data. The red rectangle  represents the signal region and the green rectangles represent the  sideband regions.  }  
\label{fig:bacground of etax}
\end{figure}

The selection criteria are also applied to the inclusive
MC sample at $\sqrt{s}=4.682$ GeV to investigate possible background contributions, which are classified into peaking and non-peaking background events.
The sources of peaking background are  listed in Table~\ref{table:The background of etax3872}, and Fig.~\ref{fig:bacground of etax} shows the $M(\gamma\gamma)$ versus $M(\pip\pim \jpsi)$ distributions for the $\ee\to\pp\psi(3823)$, $\psi(3823)\to\gamma\chi_{c1}$ background process at $\sqrt{s}=4.682$ GeV as an example. 
 The expected numbers of the peaking background events in the signal ($n^{\rm b}_{\rm sig}$) and the sideband ($n^{\rm b}_{\rm side}$) regions of $\tx$ for settings I and II are calculated by Eq.~(\ref{equ:bk number}), and   listed in  Tables~\ref{table:The expected events from different background MC samples2} and~\ref{table:The expected events from different background MC samples3}, respectively. 
The number of background events ($n^{\rm b}$) in the signal region is  corrected by 
\begin{equation}\label{equ:bk total}
\begin{split}
\begin{aligned}
    n^{\rm b}=     n^{\rm b}_{\rm non-peak}+n^{\rm b}_{\rm peak}=  
  (n^{\rm obs}_{\rm side}-n^{\rm b}_{\rm side})+   n^{\rm b}_{\rm sig},
  \end{aligned}
   \end{split}
\end{equation}   
  where   $n^{\rm obs}_{\rm side}$  is the number of observed events from the sideband regions of  $\ee\to\eta\tx$.  The number of non-peaking background  events $n^{\rm b}_{\rm non-peak}$ is represented by $n^{\rm obs}_{\rm side}-n^{\rm b}_{\rm side}$, and the number of peaking background  events $n^{\rm b}_{\rm peak}$ is  represented by $ n^{\rm b}_{\rm sig}$. 

 The  values of $n^{\rm b}_{\rm sig}$, $n^{\rm obs}_{\rm side}$, $n^{\rm b}_{\rm side}$, $n^{\rm b}$, and  the number of  observed events  in the $\eta\tx$ signal region $n^{\rm obs}_{\rm sig}$  are listed in Table~\ref{table:Preliminary results2} for  settings I and II.  
\color{black}

\begin{table}[htbp]
\renewcommand\tablename{\rm TABLE}
\caption{List of the peaking background processes~(where $\jpsi \rightarrow \LL$).}\label{table:The background of etax3872}
\begin{center}
\begin{tabular}{lll}
\hline\hline &Decay mode& \\\hline
$\EE \to \pip\pim  \psi(3823)$, &$\psi(3823) \to\gamma\chi_{c1}$,     & $\chi_{c1} \to \gamma \jpsi$ \\  
$\EE \to \omega  \chi_{cJ} (J=0,1,2)$, &$\omega \to\pip\pim\pi^{0}$,     & $\chi_{cJ} \to \gamma \jpsi$ \\  
$\EE \to \phi \chi_{cJ} (J=0,1,2)$,      &$\phi \to \pip \pim \pi^{0}$,    & $\chi_{cJ} \to \gamma \jpsi$   \\
 \hline\hline
\end{tabular}
\end{center}
\end{table}

\begin{table*}[htbp]
\renewcommand\tablename{\rm TABLE}
\caption{  Expected numbers of different peaking background events  in  $\ee\to\eta\tx$ signal and sideband regions at different energy points for setting I, where $n^{\rm b}_{\rm sig}$ and $n^{\rm b}_{\rm side}$ are the total sums for the signal and the sideband regions, respectively. Ellipses represent results close to 0 or 0. }
 \label{table:The expected events from different background MC samples2}
\begin{center}
\begin{tabular}{ccccccccc}
\hline\hline   
\multirow{13}{*}{Signal region}&$\sqrt s$ (GeV)&   $\phi\chi_{c1}$ &  $\phi\chi_{c2}$ &   $\omega\chi_{c1}$ & $\omega\chi_{c2}$        &$\pip\pim\psi(3823)$ &$n^{\rm b}_{\rm sig}$\\\hline
    &4.612   &$\cdots$&$\cdots$&$0.02   \pm0.03  $&$0.02$&$0.11 \pm0.14    $&$0.15   \pm0.11$\\     &4.628   &$0.01$&$0.01$&$0.08   \pm0.03  $&$0.03$&$0.48 \pm0.26    $&$0.61   \pm0.49$\\ 
    &4.641   &$0.01$&$0.01$&$0.10   \pm0.04  $&$0.04$&$0.57 \pm0.22    $&$0.73   \pm0.58$\\     &4.661   &$\cdots$&$0.01$&$0.07   \pm0.03  $&$0.05  \pm0.03  $&$0.54 \pm0.17    $&$0.69   \pm0.55$\\   
    &4.682   &$0.01$&$0.03   \pm0.01   $&$0.05   \pm0.04  $&$0.04  \pm0.05  $&$0.64 \pm0.17    $&$0.77   \pm0.64$\\    &4.699   &$\cdots$&$0.01$&$0.03   \pm0.01  $&$0.02$&$0.07 \pm0.04    $&$0.14   \pm0.08$\\ 
    &4.740   &$\cdots$&$\cdots$&$0.00   \pm0.01  $&$0.01$&$0.01 \pm0.01    $&$0.02   \pm0.01$\\     &4.750   &$\cdots$&$\cdots$&$0.01$&$0.01$&$0.01 \pm0.01    $&$0.04   \pm0.02$\\ 
    &4.780   &$\cdots$&$\cdots$&$0.00   \pm0.01  $&$0.01  \pm0.01  $&$0.01$&$0.02   \pm0.01$\\      &4.843   &$\cdots$&$\cdots$&$\cdots$&$0.00  \pm0.01  $&$\cdots$&$0.01   \pm0.01$\\
    &4.918   &$\cdots$&$\cdots$&$\cdots$&$\cdots$&$\cdots$&$0.01$\\   
\hline
\multirow{13}{*}{Sideband region}&$\sqrt s$ (GeV)&    $\phi\chi_{c1}$ &  $\phi\chi_{c2}$ &   $\omega\chi_{c1}$ & $\omega\chi_{c2}$          &$\pip\pim\psi(3823)$ &$n^{\rm b}_{\rm side}$\\\hline
     &4.612  &$\cdots$&$\cdots$&$0.01      \pm0.02      $&$0.01    $&$0.01      \pm0.02          $&$0.03         \pm0.02$\\         &4.628  &$0.01$&$\cdots$&$0.04      \pm0.02      $&$0.02 $&$0.14      \pm0.08          $&$0.22         \pm0.15$\\        
    &4.641  &$0.01$&$\cdots$&$0.06      \pm0.02      $&$0.02 $&$0.28      \pm0.11          $&$0.37         \pm0.29$\\            &4.661  &$\cdots$&$0.01$&$0.04      \pm0.02      $&$0.03      \pm0.02      $&$0.51      \pm0.16          $&$0.60         \pm0.51$\\         
    &4.682  &$0.01$&$0.03      \pm0.01      $&$0.04      \pm0.03      $&$0.03      \pm0.04      $&$1.00      \pm0.27          $&$1.12         \pm1.00$\\             &4.699  &$\cdots$&$0.02$&$0.03      \pm0.01      $&$0.03   $&$0.16      \pm0.09          $&$0.24         \pm0.17$\\      
    &4.740  &$\cdots$&$\cdots$&$0.01      \pm0.01      $&$0.01   $&$0.05      \pm0.07          $&$0.07         \pm0.05$\\          &4.750  &$\cdots$&$\cdots$&$0.01$&$0.01 $&$0.16      \pm0.09          $&$0.19         \pm0.16$\\        
    &4.780  &$\cdots$&$\cdots$&$0.01      \pm0.01      $&$0.01      \pm0.02      $&$0.10      \pm0.06          $&$0.12         \pm0.10$\\    &4.843  &$\cdots$&$\cdots$&$\cdots$&$0.00      \pm0.01      $&$0.00      \pm0.01          $&$0.01         \pm0.01$\\ 
    &4.918  &$\cdots$&$\cdots$&$\cdots$&$\cdots$&$\cdots$&$0.01$\\  
 \hline\hline
\end{tabular}
\end{center}
\end{table*}

\begin{table*}[htbp]
\renewcommand\tablename{\rm TABLE}
\caption{ Expected numbers of different peaking background events in  $\ee\to\eta\tx$ signal  and sideband regions at different energy points for setting II, where $n^{\rm b}_{\rm sig}$ and $n^{\rm b}_{\rm side}$ are the total sums for the signal and the sideband regions, respectively. Ellipses represent results close to 0 or 0.
}\label{table:The expected events from different background MC samples3}
\begin{center}
\begin{tabular}{ccccccc}
\hline\hline   
\multirow{13}{*}{Signal region}&$\sqrt s$ (GeV)&   $\phi\chi_{c2}$ &   $\omega\chi_{c1}$ & $\omega\chi_{c2}$        &$\pip\pim\psi(3823)$ &$n^{\rm b}_{\rm sig}$\\\hline
&4.612       &$\cdots$&$0.00      \pm0.01      $&$\cdots$&$0.02      \pm0.03         $&$0.03         \pm0.03$\\
&4.628       &$\cdots$&$0.02      \pm0.01      $&$0.01$&$0.10      \pm0.06         $&$0.13         \pm0.11$\\
&4.641       &$\cdots$&$0.02      \pm0.01      $&$0.01$&$0.12      \pm0.05         $&$0.16         \pm0.12$\\
&4.661       &$\cdots$&$0.02      \pm0.01      $&$0.01      \pm0.01      $&$0.10      \pm0.03         $&$0.13         \pm0.10$\\
&4.682       &$\cdots$&$0.01      \pm0.01      $&$0.01      \pm0.01      $&$0.09      \pm0.02         $&$0.12         \pm0.09$\\
&4.699       &$\cdots$&$0.01$&$\cdots$&$0.01$&$0.02         \pm0.01$\\
&4.740       &$\cdots$&$\cdots$&$\cdots$&$\cdots$&$\cdots$\\
&4.750       &$\cdots$&$\cdots$&$\cdots$&$\cdots$&$0.01$\\
\hline
\multirow{13}{*}{Sideband region}&$\sqrt s$ (GeV)&  $\phi\chi_{c2}$ &   $\omega\chi_{c1}$ & $\omega\chi_{c2}$          &$\pip\pim\psi(3823)$ &$n^{\rm b}_{\rm side}$\\\hline
&4.612       &$\cdots$&$0.00      \pm0.01      $&$\cdots$&$0.02      \pm0.03         $&$0.03         \pm0.02$\\&4.628       &$\cdots$&$0.01      \pm0.01      $&$0.01$&$0.10      \pm0.05         $&$0.12         \pm0.10$\\
&4.641       &$\cdots$&$0.02      \pm0.01      $&$0.01$&$0.11      \pm0.04         $&$0.14         \pm0.11$\\&4.661       &$\cdots$&$0.02      \pm0.01      $&$0.01      \pm0.01      $&$0.10      \pm0.03         $&$0.13         \pm0.11$\\
&4.682       &$0.01$&$\cdots$&$0.01      \pm0.01      $&$0.10      \pm0.03         $&$0.13         \pm0.10$\\&4.699       &$\cdots$&$0.01$&$\cdots$&$0.01      \pm0.01         $&$0.03         \pm0.01$\\
&4.740       &$\cdots$&$\cdots$&$\cdots$&$\cdots$&$\cdots$\\&4.750       &$\cdots$&$\cdots$&$\cdots$&$\cdots$&$0.01$\\
 \hline\hline
\end{tabular}
\end{center}
\end{table*}

\subsection{Cross section measurement}
\label{sec:cross2}
The signal yield of $\ee\to\eta\tx$, $n^{\rm sig}$, is obtained by $n^{\rm sig} = {\rm max}(0, n^{\rm obs}_{\rm sig}- n^{\rm b})$ according to  the  F-C method~\cite{feldman-cousins-method}, and the upper limit
  is provided at   90\% C.L.\ by  the POLE method since no significant  signal is observed.  The statistical significance at each energy point is calculated  using the $P$-value method and is summarized  in Table~\ref{table:Preliminary results2}. 

 The product of  the $\ee\to\eta\tx$ cross section and the $\tx\to\ppjpsi$ branching fraction, together with the upper limit at 90\% C.L.,  are calculated by  
\begin{equation}
\begin{split}
\begin{aligned}
 \sigma^{\rm B}[\ee\to\eta\tx]B[\tx\to\ppjpsi]= \\
 \frac{n^{\rm sig}}
 {\mathcal{L_{\rm int}}(1+\delta)|1-\Pi|^{-2}
   \mathcal{B}_2
  (\epsilon_e\mathcal{B}_e+\epsilon_\mu\mathcal{B}_\mu)
   },
  \end{aligned}
   \end{split}
\end{equation}
where   $(1+\delta)$ is the radiative correction factor obtained from  the quantum electrodynamics calculation~\cite{isr-calculate2} using the {\sc kkmc} generator~\cite{ref-kkmc, ref-kkmc-2}, assuming that the c.m.\ energy dependence cross section follows the  line shape of  the power function $1/s$.

\begin{figure}[htbp]
\renewcommand\figurename{\rm FIG}
\begin{center} 
 \includegraphics[width=3.0in,height=2.0in]
 {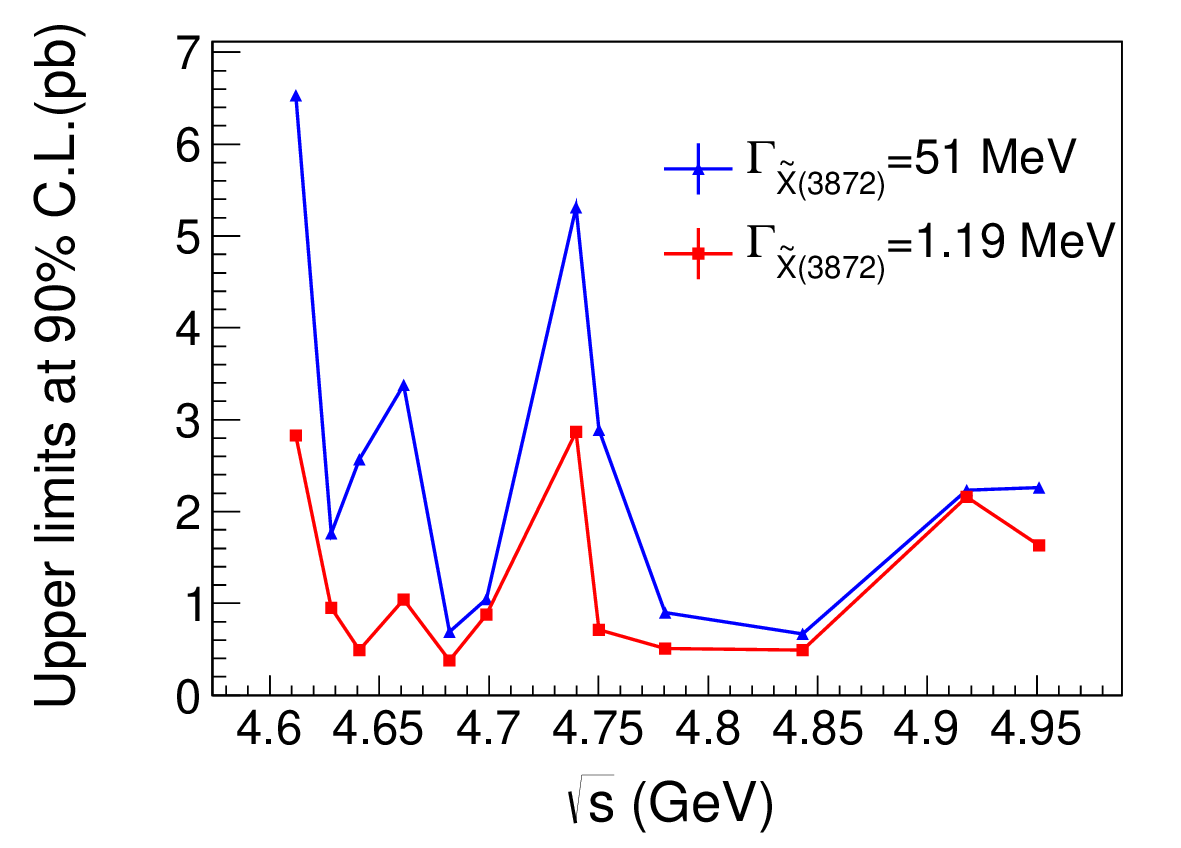}
\end{center}
\caption{Upper limits of $\sigma^{\rm B}[\ee\to\eta\tx]B[\tx\to\ppjpsi]$  at 90\% C.L. for different assumptions of the $\tx$ widths.
 }   
\label{fig:cross section2}
\end{figure}

The  $\sigma^{\rm B}[\ee\to\eta\tx]B[\tx\to\ppjpsi]$ values,  the corresponding  upper limits at 90\% C.L.,  together with the  other information used for the cross section calculation, are listed in 
Table~\ref{table:Preliminary results2}, separately for settings I and II.  The $\sigma^{\rm B}[\ee\to\eta\tx]B[\tx\to\ppjpsi]$ upper limits at 90\% C.L.\ for the two settings are shown in Fig.~\ref{fig:cross section2}.

\begin{sidewaystable*}[htbp]
\renewcommand\tablename{\rm TABLE}
\vspace{8cm}
\caption{The $\sigma^{\rm B}[\ee\to\eta\tx]B[\tx\to\ppjpsi]$ ($\sigma B$) values and the corresponding upper limits   at   90\% C.L.\ (UL)
with the POLE  method for settings I and II at different energy points, together with the integrated luminosity
$\mathcal{L}_{\rm int}$, the numbers of observed events  in signal $n^{\rm obs}_{\rm sig}$   and sideband $n^{\rm obs}_{\rm side}$ regions, the  expected numbers of peaking background events in signal $n^{\rm b}_{\rm sig}$ and sideband $n^{\rm b}_{\rm side}$ regions, the total yield  of background events in signal region $n^{\rm b}=n^{\rm obs}_{\rm side}+n^{\rm b}_{\rm sig}-n^{\rm b}_{\rm side}$,  
the numbers of signal events $n^{\rm sig}$, the upper limits
at  90\% C.L.\  of the number of signal events $n^{\rm sig}_{\rm POLE}$, 
the products of detection efficiencies and branching fractions
$\Sigma=\mathcal{B}_2(\epsilon_e\mathcal{B}_e+\epsilon_\mu\mathcal{B}_\mu)$,
the products of  ISR correction factor  and  vacuum polarization factor $(1+\delta)|1-\Pi|^{-2}$, and the signal statistical significances.
The uncertainties of $n^{\rm sig}$ and $\sigma B$ are statistical only. The significance is calculated  using the $P$-value method.
}\label{table:Preliminary results2}
\begin{center}
\begin{tabular}{ccccccccccccccc}
\hline\hline
\multirow{13}{*}{Setting I}&$\sqrt s$~(GeV) &$ \mathcal{L}_{\rm int}$~(pb$^{-1}$)&$n^{\rm obs}_{\rm sig}$&$n^{\rm obs}_{\rm side}$&$n^{\rm b}_{\rm sig}$&$n^{\rm b}_{\rm side}$&$n^{\rm b}$ &$n^{\rm sig}$&$n^{\rm sig}_{\rm POLE}$
& $\Sigma(10^{-2})$ &$(1+\delta)|1-\Pi|^{-2}$ & $\sigma B$~(pb)  & UL~(pb)&significance\\\hline
     &4.612&    103.7&  1&  0&0.13&0.03&  0.11&              $0.9_{-0.6}^{+1.8}$&           (0.0,~4.2)&    0.848&  0.73&                   $1.4_{-1.0}^{+2.7}$&           (0.0,~6.5)&$1.2\sigma$\\
    &4.628&    521.5&  3&  1&0.62&0.22&  1.40&              $1.6_{-1.2}^{+2.3}$&           (0.0,~6.0)&    0.835&  0.79&                   $0.5_{-0.4}^{+0.7}$&           (0.0,~1.8)&$1.0\sigma$\\
    &4.641&    551.7&  5&  0&0.72&0.37&  0.35&              $4.7_{-2.2}^{+2.8}$&           (1.6,~9.6)&    0.828&  0.81&                   $1.3_{-0.6}^{+0.8}$&           (0.4,~2.6)&$4.0\sigma$\\
    &4.661&    529.4&  7&  1&0.69&0.60&  1.08&              $5.9_{-2.8}^{+3.3}$&           (2.4,~11.8)&    0.788&  0.84&                   $1.7_{-0.8}^{+0.9}$&          (0.7,~3.4)&$3.6\sigma$\\
    &4.682&   1667.4&  7&  6&0.82&1.16&  5.66&              $1.3_{-1.3}^{+3.3}$&           (0.0,~7.6)&    0.765&  0.86&                   $0.1_{-0.1}^{+0.3}$&           (0.0,~0.7)&$0.4\sigma$\\
    &4.699&    535.5&  2&  3&0.12&0.21&  2.91&              $0.0_{-0.0}^{+1.5}$&           (0.0,~3.7)&    0.757&  0.87&                   $0.0_{-0.0}^{+0.4}$&           (0.0,~1.1)&$\cdots$\\
    &4.740&    163.9&  2&  0&0.01&0.05&  0.00&              $2.0_{-1.3}^{+2.3}$&           (0.5,~5.8)&    0.750&  0.89&                   $1.8_{-1.2}^{+2.1}$&           (0.5,~5.3)&$\cdots$\\
    &4.750&    366.5&  4&  2&0.02&0.17&  1.85&              $2.2_{-1.6}^{+2.8}$&           (0.0,~7.1)&    0.743&  0.90&                   $0.9_{-0.6}^{+1.1}$&           (0.0,~2.9)&$1.2\sigma$\\
    &4.781&    511.5&  1&  3&0.02&0.12&  2.90&              $0.0_{-0.0}^{+0.5}$&           (0.0,~3.1)&    0.738&  0.91&                   $0.0_{-0.0}^{+0.1}$&           (0.0,~0.9)&$\cdots$\\
    &4.843&    525.2&  0&  3&0.00&0.01&  2.99&              $0.0_{-0.0}^{+0.1}$&           (0.0,~2.4)&    0.726&  0.93&                   $0.0_{-0.0}^{+0.0}$&           (0.0,~0.7)&$\cdots$\\
    &4.918&    207.8&  1&  2&0.00&0.00&  2.00&              $0.0_{-0.0}^{+1.0}$&           (0.0,~3.2)&    0.714&  0.95&                   $0.0_{-0.0}^{+0.7}$&           (0.0,~2.2)&$\cdots$\\
    &4.951&    159.3&  0&  0&0.00&0.00&  0.00&              $0.0_{-0.0}^{+1.3}$&           (0.0,~2.4)&    0.708&  0.96&                   $0.0_{-0.0}^{+1.2}$&           (0.0,~2.3)&$\cdots$\\  
\hline 
\multirow{13}{*}{Setting II}&$\sqrt s$~(GeV) &$ \mathcal{L}_{\rm int}$~(pb$^{-1}$)&$n^{\rm obs}_{\rm sig}$&$n^{\rm obs}_{\rm side}$&$n^{\rm b}_{\rm sig}$&$n^{\rm b}_{\rm side}$&$n^{\rm b}$ &$n^{\rm sig}$&$n^{\rm sig}_{\rm POLE}$
& $\Sigma(10^{-2})$ &$(1+\delta)|1-\Pi|^{-2}$ & $\sigma B$~(pb)  & UL~(pb)&significance\\\hline
      &4.612&    103.7&  0&  0&0.03&0.03&  0.00&              $0.0_{-0.0}^{+1.3}$&           (0.0,~2.4)&    1.136&  0.73&                   $0.0_{-0.0}^{+1.5}$&           (0.0,~2.8)&$\cdots$\\
    &4.628&    521.5&  1&  0&0.14&0.13&  0.01&              $1.0_{-0.6}^{+1.8}$&           (0.0,~4.4)&    1.119&  0.79&                   $0.2_{-0.1}^{+0.4}$&           (0.0,~1.0)&$2.4\sigma$\\
    &4.641&    551.7&  0&  1&0.15&0.14&  1.02&              $0.0_{-0.0}^{+0.5}$&           (0.0,~2.4)&    1.110&  0.81&                   $0.0_{-0.0}^{+0.1}$&           (0.0,~0.5)&$\cdots$\\
    &4.661&    529.4&  2&  1&0.13&0.14&  1.00&              $1.0_{-0.8}^{+2.3}$&           (0.0,~5.0)&    1.083&  0.84&                   $0.2_{-0.2}^{+0.5}$&           (0.0,~1.0)&$0.6\sigma$\\
    &4.682&   1667.4&  2&  0&0.13&0.13&  0.00&              $2.0_{-1.3}^{+2.3}$&           (0.5,~5.8)&    1.063&  0.86&                   $0.1_{-0.1}^{+0.1}$&           (0.0,~0.4)&$\cdots$\\
    &4.699&    535.5&  1&  0&0.02&0.02&  0.00&              $1.0_{-0.6}^{+1.8}$&           (0.0,~4.3)&    1.048&  0.87&                   $0.2_{-0.1}^{+0.4}$&           (0.0,~0.9)&$\cdots$\\
    &4.740&    163.9&  1&  0&0.00&0.00&  0.00&              $1.0_{-0.6}^{+1.8}$&           (0.0,~4.4)&    1.044&  0.89&                   $0.7_{-0.4}^{+1.2}$&           (0.0,~2.9)&$\cdots$\\
    &4.750&    366.5&  0&  1&0.00&0.00&  1.00&              $0.0_{-0.0}^{+0.5}$&           (0.0,~2.4)&    1.034&  0.90&                   $0.0_{-0.0}^{+0.2}$&           (0.0,~0.7)&$\cdots$\\
    &4.781&    511.5&  0&  0&0.00&0.00&  0.00&              $0.0_{-0.0}^{+1.3}$&           (0.0,~2.4)&    1.032&  0.91&                   $0.0_{-0.0}^{+0.3}$&           (0.0,~0.5)&$\cdots$\\
    &4.843&    525.2&  0&  0&0.00&0.00&  0.00&              $0.0_{-0.0}^{+1.3}$&           (0.0,~2.4)&    1.018&  0.93&                   $0.0_{-0.0}^{+0.3}$&           (0.0,~0.5)&$\cdots$\\
    &4.918&    207.8&  1&  0&0.00&0.00&  0.00&              $1.0_{-0.6}^{+1.8}$&           (0.0,~4.3)&    0.999&  0.95&                   $0.5_{-0.3}^{+0.9}$&           (0.0,~2.2)&$3.1\sigma$\\
    &4.951&    159.3&  0&  0&0.00&0.00&  0.00&              $0.0_{-0.0}^{+1.3}$&           (0.0,~2.4)&    0.981&  0.96&                   $0.0_{-0.0}^{+0.9}$&           (0.0,~1.6)&$\cdots$\\
\hline\hline
\end{tabular}
\end{center}
\end{sidewaystable*}

\color{black}

\section{\boldmath Systematic uncertainties}
\label{sec:uncertainty}
The systematic uncertainties for the cross-section measurements of the $\ee\to\eta\psip$ and $\ee\to\eta\tx$[$\tx\to\ppjpsi$] processes are mainly due to the integrated  luminosity, the tracking and photon reconstruction,   the branching fractions of intermediate particle decays, the ISR correction factor, the kinematic fit, the background estimation, and the mass windows. These  systematic uncertainties are explained below and summarized in Tables~\ref{table:Unscertainties} and~\ref{table:Unscertainties2}. The total systematic uncertainty is obtained by summing the individual uncertainties in quadrature, assuming that all sources are independent.

\begin{itemize}
\item \emph{Luminosity.} The uncertainty of integrated luminosity is $1.0\%$, measured using Bhabha scattering events~\cite{luminosity-measurement, luminosity-measurement-2}.

\item \emph{Tracking efficiency.} The uncertainty of the tracking efficiency is 1.0\% per single track from Ref.~\cite{intro-bes3-etapjpsi}. Since there are four charged tracks in the $\ee\to\eta\psip$ and $\eta\tx$ final states, the total uncertainty due to tracking efficiency is 4.0\%.

\item \emph{Photon  detection efficiency.} The uncertainty from photon reconstruction is $1.0\%$ per photon, from the study of the process $\jpsi\to\rho^{0}\piz$, $\rho^{0}\to\pp$, $\piz\to\gamma\gamma$~\cite{photoneffi}. Thus the total uncertainty due to photon reconstruction is 2.0\%.  

\item \emph{Branching fraction.} The uncertainties on the branching fractions of the intermediate states from $\ee\to\eta\psip$ and $\ee\to\eta\tx\to\eta\ppjpsi$ are 1.2\% and 0.8\%, respectively, from the  PDG values~\cite{pdg}. 

\item \emph{ISR correction factor.} The ISR factors are  obtained from the input line shapes of the $Y(4260)$ state at the energy points below 4.600 GeV and the power function $1/s$  at the energy points above 4.600 GeV.  To estimate the uncertainties, the input line shapes are replaced by the  $\psi(4415)$ state at the energy points below 4.600 GeV and the $Y(4660)$ states at the energy points above 4.600 GeV; the $(1+\delta)\epsilon$ differences for the different assumptions are taken as the systematic uncertainties, and they range from 0.2\% to 18.3\%. 

\item \emph{Kinematic fit.} The systematic uncertainty from the kinematic fit is estimated by correcting the helix parameters of charged tracks according to the method described in Ref.~\cite{bes3-kinematicfit-eff}. The MC sample with the track helix parameter correction is considered as the nominal one. The difference between detection efficiencies
obtained from MC samples with and without the correction is taken as the uncertainty. They range from 1.4\% to 3.8\%.

\item \emph{Background estimation.}
The dominant uncertainty for the number of $\ee\to\gamma\gamma\psip$ background events is statistical only, and it is measured using Eq.~(\ref{equ:f factor}).
For the other background sources, the numbers of events  
are estimated using the   Eq.~(\ref{equ:bk number}). The  cross sections mean values for the background are increased by their corresponding uncertainties, resulting in a change of the signal yields. The difference of the signal cross sections due to this change is taken as the uncertainty of the background estimation. The uncertainties range from 0.0\% to 45.0\% for the $\ee\to\eta\psip$ and from  0.0\% to 16.9\% for the $\ee\to\eta\tx$ cross section measurements. For some cases, the  change does not affect the number of signal events, thus their uncertainty due to background estimation shows zero in Tables~\ref{table:Unscertainties} and \ref{table:Unscertainties2}.

\item \emph{Mass window.} The systematic uncertainties on  $\jpsi$,  $\eta$, and $\psip$ mass windows are  0.9\%, 2.6\%, and 2.3\%, respectively, from Ref.~\cite{etapsip}. The total systematic uncertainty on mass windows for the $\ee\to\eta\psip$ process is 3.6\% by adding the three numbers in quadrature, while for the $\ee\to\eta\tx$ process is 2.8\% summing up quadratically the uncertainties of the $\jpsi$ and $\eta$ mass windows.

\item \emph{Sideband region.} The systematic uncertainty due to the sideband region for  the $\ee\to\eta\tx$ cross section measurement is estimated by moving the region of $\pm0.5\sigma$ of the $\tx$ mass. The difference of cross sections is taken as the systematic uncertainty on sideband region. 
 
\end{itemize}

\begin{table*}[htbp]
\renewcommand\tablename{\rm TABLE}
\caption{The  systematic uncertainties of $\ee\to\eta\psi(2S)$ cross section measurement for luminosity (Lum), tracking efficiency,  photon reconstruction efficiency,  branching fraction (BR), ISR  correction factor, kinematic fit (KMFIT), background estimation (BK), and mass windows (MW)~(in units of \%). Ellipses mean that the results are  0.}\label{table:Unscertainties}
\begin{center} 
\begin{tabular}{cccccccccc}
\hline\hline
$\sqrt s$~(GeV) & Lum& Tracking& Photon&BR&MW&KMFIT&ISR&BK&Total\\\hline
  4.288 & 1.0  & 4.0 &  2.0  & 1.2  & 3.6 &  3.3 &  3.6 & 19.5 & 21.0\\     
  4.312 & 1.0  & 4.0 &  2.0  & 1.2  & 3.6 &  3.1 &  9.7 & 30.6 & 32.8\\     
  4.337 & 1.0  & 4.0 &  2.0  & 1.2  & 3.6 &  3.8 & 15.6 & 45.0 & 48.2\\     
  4.377 & 1.0  & 4.0 &  2.0  & 1.2  & 3.6 &  3.3 & 18.3 & 10.7 & 22.3\\     
  4.396 & 1.0  & 4.0 &  2.0  & 1.2  & 3.6 &  3.3 & 16.9 &  $\cdots$ & 18.2\\     
  4.436 & 1.0  & 4.0 &  2.0  & 1.2  & 3.6 &  3.2 & 13.0 & 10.5 & 18.0\\     
  4.612 &  1.0  & 4.0 &  2.0  & 1.2  & 3.6 &  2.7 &  5.7 &  $\cdots$ &  8.7\\ 
  4.628 &  1.0  & 4.0 &  2.0  & 1.2  & 3.6 &  3.0 &  4.2 &  $\cdots$ &  7.9\\
  4.641 &  1.0  & 4.0 &  2.0  & 1.2  & 3.6 &  3.0 &  1.9 & 20.1 & 21.3\\ 
  4.661 &  1.0  & 4.0 &  2.0  & 1.2  & 3.6 &  2.9 &  7.4 & 25.9 & 27.8\\
  4.682 &  1.0  & 4.0 &  2.0  & 1.2  & 3.6 &  2.7 & 11.6 & $\cdots$ & 13.3\\ 
  4.699 &  1.0  & 4.0 &  2.0  & 1.2  & 3.6 &  2.7 & 11.4 & 40.6 & 42.7\\
  4.740 &  1.0  & 4.0 &  2.0  & 1.2  & 3.6 &  2.9 &  8.3 &  $\cdots$ & 10.7\\ 
  4.750 &  1.0  & 4.0 &  2.0  & 1.2  & 3.6 &  2.7 &  8.4 & 13.5 & 17.2\\
  4.781 &  1.0  & 4.0 &  2.0  & 1.2  & 3.6 &  2.5 &  6.4 & $\cdots$ &  9.1\\ 
  4.843 &  1.0  & 4.0 &  2.0  & 1.2  & 3.6 &  2.9 &  3.7 & 44.4 & 45.1\\
  4.918 &  1.0  & 4.0 &  2.0  & 1.2  & 3.6 &  2.7 &  4.5 &  $\cdots$ &  8.0\\ 
  4.951 &  1.0  & 4.0 &  2.0  & 1.2  & 3.6 &  3.0 &  3.3 &  $\cdots$ &  7.4\\
    \hline\hline
    \end{tabular}
  \end{center}
\end{table*}

\begin{table*}[htbp]
\renewcommand\tablename{\rm TABLE}
\caption{The  systematic uncertainties of $\ee\to\eta\tx$ cross section measurement for luminosity (Lum),  tracking efficiency,  photon reconstruction efficiency,  branching fraction (BR), $\eta$ and $\jpsi$ mass windows (MW), kinematic fit (KMFIT),  ISR  correction factor, background expectation (BK),  and  sideband regions (Sidebands) for settings I and II~(in units of \%). The total  systematic uncertainty  is the sum of the individual uncertainties in quadrature. Ellipses mean that the results are  0.   
}\label{table:Unscertainties2}
\begin{center}
\begin{tabular}{cccccccccccc}
\hline\hline
\multirow{13}{*}{Setting I}&$\sqrt s$~(GeV) & Lum & Tracking& Photon&BR &MW&KMFIT&ISR&BK&Sidebands& Total\\\hline
 &4.612 & 1.0  & 4.0 &  2.0  & 0.8 &  2.8 &  1.4 &  3.1 & 16.9 &  0.1 & 18.1\\
 &4.628 & 1.0  & 4.0 &  2.0  & 0.8 &  2.8 &  1.6 &  1.7 & 13.8 &  0.1 & 15.0\\
 &4.641 & 1.0  & 4.0 &  2.0  & 0.8 &  2.8 &  1.9 &  1.2 &  3.2 &  0.1 &  6.7\\
 &4.661 & 1.0  & 4.0 &  2.0  & 0.8 &  2.8 &  1.9 &  7.8 &  0.7 &  0.1 &  9.7\\
 &4.682 & 1.0  & 4.0 &  2.0  & 0.8 &  2.8 &  1.9 & 11.8 &  6.0 &  0.1 & 14.4\\
 &4.699 & 1.0  & 4.0 &  2.0  & 0.8 &  2.8 &  1.5 &  9.9 &  $\cdots$ &  0.1 & 11.4\\
 &4.740 & 1.0  & 4.0 &  2.0  & 0.8 &  2.8 &  1.7 &  7.7 &  $\cdots$ &  0.1 &  9.6\\
 &4.750 & 1.0  & 4.0 &  2.0  & 0.8 &  2.8 &  1.6 &  8.4 &  3.7 &  0.1 & 10.8\\
 &4.781 & 1.0  & 4.0 &  2.0  & 0.8 &  2.8 &  1.9 &  6.4 &  $\cdots$ &  0.1 &  8.6\\
 &4.843 & 1.0  & 4.0 &  2.0  & 0.8 &  2.8 &  1.4 &  4.8 &  $\cdots$ &  0.1 &  7.4\\
 &4.918 & 1.0  & 4.0 &  2.0  & 0.8 &  2.8 &  1.5 &  3.4 &  $\cdots$ &  0.1 &  6.6\\
 &4.951 & 1.0  & 4.0 &  2.0  & 0.8 &  2.8 &  1.7 &  5.6 &  $\cdots$ &  0.1 &  8.0\\
\hline 
\multirow{13}{*}{Setting II}&$\sqrt s$~(GeV) & Lum & Tracking& Photon&BR &MW&Kine-fit&ISR&BK&Sidebands &Total\\\hline
   &4.612 &  1.0  & 4.0 &  2.0 &  0.8 &  2.8  & 1.8 &  3.3 & $\cdots$ & 1.3 &  6.7\\ 
   &4.628 &  1.0  & 4.0 &  2.0 &  0.8 &  2.8  & 1.8 &  1.6 & $\cdots$ & 1.3 &  6.1\\ 
   &4.641 &  1.0  & 4.0 &  2.0 &  0.8 &  2.8  & 2.3 &  0.2 & $\cdots$ & 1.3 &  6.1\\ 
   &4.661 &  1.0  & 4.0 &  2.0 &  0.8 &  2.8  & 2.3 &  6.4 & $\cdots$ & 1.3 &  8.8\\ 
   &4.682 &  1.0  & 4.0 &  2.0 &  0.8 &  2.8  & 2.2 & 10.3 & $\cdots$ & 1.3 & 12.0\\
   &4.699 &  1.0  & 4.0 &  2.0 &  0.8 &  2.8  & 2.2 &  9.8 & $\cdots$ & 1.3 & 11.5\\
   &4.740 &  1.0  & 4.0 &  2.0 &  0.8 &  2.8  & 2.2 &  7.3 & $\cdots$ & 1.3 &  9.5\\ 
   &4.750 &  1.0  & 4.0 &  2.0 &  0.8 &  2.8  & 2.0 &  6.4 & $\cdots$ & 1.3 &  8.8\\ 
   &4.781 &  1.0  & 4.0 &  2.0 &  0.8 &  2.8  & 2.1 &  6.3 & $\cdots$ & 1.3 &  8.7\\ 
   &4.843 &  1.0  & 4.0 &  2.0 &  0.8 &  2.8  & 2.1 &  4.3 & $\cdots$ & 1.3 &  7.4\\ 
   &4.918 &  1.0  & 4.0 &  2.0 &  0.8 &  2.8  & 2.1 &  3.3 & $\cdots$ & 1.3 &  6.8\\ 
   &4.951 &  1.0  & 4.0 &  2.0 &  0.8 &  2.8  & 2.2 &  4.8 & $\cdots$ & 1.3 &  7.7\\ 
\hline\hline
    \end{tabular}
\end{center}
\end{table*}

\section{\boldmath Summary}
\label{sec:summ}
 
In summary, using the new BESIII data samples collected at c.m.\ energies from 4.288  GeV to 4.951 GeV with an integrated  luminosity of about 8.9 $\rm fb^{-1}$,   
the process $\EE\to \eta\psi(2S)$ with $\psip\to\ppjpsi$ has been  studied. 
Due to the  limited statistics,   upper limits on the cross sections are provided at  90\% C.L.   
Combining the current measurement and  the previous result from Ref.~\cite{etapsip}, the total number of observed events at the 32 energy points  is 79, and the total number of expected background events  is 27,  
 corresponding to a  $P$-value of  $5.9\times10^{-16}$  for the background-only hypothesis and a statistical significance of  more than $5\sigma$ for $\ee\to \eta\psip$  based on $14.2~{\rm fb}^{-1}$  of BESIII data.

In addition, we have searched for signals of the $\tx$   with charge conjugate parity $C=-1$ and a mass of $3860.0\pm10.4~\mevcc$, as reported by the  COMPASS experiment~\cite{compass}  in the $\ppjpsi$ invariant mass distribution.  
We do not find any evident signal for the $\tx$, and we set upper limits  on the cross sections of $\ee\to\eta\tx$[$\tx\to\ppjpsi$]  at 90\% C.L.\ assuming two different width hypotheses, which are consistent with each other.  
 \\

 \textbf{Acknowledgement}

The BESIII Collaboration thanks the staff of BEPCII and the IHEP computing center for their strong support. This work is supported in part by National Key R\&D Program of China under Contracts Nos. 2020YFA0406300, 2020YFA0406400, 2023YFA1606000; National Natural Science Foundation of China (NSFC) under Contracts Nos. 11635010, 11735014, 11835012, 11935015, 11935016, 11935018, 11961141012, 12025502, 12035009, 12035013, 12061131003, 12192260, 12192261, 12192262, 12192263, 12192264, 12192265, 12221005, 12225509, 12235017; the Chinese Academy of Sciences (CAS) Large-Scale Scientific Facility Program; the CAS Center for Excellence in Particle Physics (CCEPP); Joint Large-Scale Scientific Facility Funds of the NSFC and CAS under Contract No. U1832207; CAS Key Research Program of Frontier Sciences under Contracts Nos. QYZDJ-SSW-SLH003, QYZDJ-SSW-SLH040; 100 Talents Program of CAS; The Institute of Nuclear and Particle Physics (INPAC) and Shanghai Key Laboratory for Particle Physics and Cosmology; European Union's Horizon 2020 research and innovation programme under Marie Sklodowska-Curie grant agreement under Contract No. 894790; German Research Foundation DFG under Contracts Nos. 455635585, Collaborative Research Center CRC 1044, FOR5327, GRK 2149; Istituto Nazionale di Fisica Nucleare, Italy; Ministry of Development of Turkey under Contract No. DPT2006K-120470; National Research Foundation of Korea under Contract No. NRF-2022R1A2C1092335; National Science and Technology fund of Mongolia; National Science Research and Innovation Fund (NSRF) via the Program Management Unit for Human Resources \& Institutional Development, Research and Innovation of Thailand under Contract No. B16F640076; Polish National Science Centre under Contract No. 2019/35/O/ST2/02907; The Swedish Research Council; U. S. Department of Energy under Contract No. DE-FG02-05ER41374


\begin{thebibliography}{99}


\bibitem{x3872} S. K. Choi {\it et al.} (Belle Collaboration), Observation of a narrow charmonium-like state in exclusive $B\pm\to K^{\pm}\ppjpsi$  decays,  
 \href{http://dx.doi.org/10.1103/PhysRevLett.91.262001}{Phys.\ Rev.\ Lett.\ {\bf 91}, 262001 (2003)}.

\bibitem{theory-Y-states-Brambilla-2020} N. Brambilla, S. Eidelman, C. Hanhart, A. Nefediev,
C. P. Shen, C. E. Thomas, A. Vairo, and C. Z. Yuan,  The XYZ states: Experimental and theoretical status and perspectives, 
 \href{https://www.sciencedirect.com/science/article/pii/S0370157320301915}{  Phys.\ Rept.\ {\bf 873}, 1 (2020)}.
 
\bibitem{zcs3985}  M. Ablikim {\it et al.} (BESIII Collaboration),  Observation of a Near-Threshold Structure in the $K^{+}$  Recoil-Mass Spectra in $\ee\to K^{+}(D_s^{-}D^{*0}+D_s^{*-}D^0)$,  \href{https://journals.aps.org/prl/abstract/10.1103/PhysRevLett.126.102001}{Phys.\ Rev.\ Lett.\ {\bf 126}, 102001 (2021)}.

\bibitem{chao2} T. Barnes, S. Godfrey, and E. S. Swanson,  Higher charmonia,
 \href{https://journals.aps.org/prd/abstract/10.1103/PhysRevD.72.054026}{ Phys.\ Rev.\ D {\bf 72},  054026 (2005)}.

    \bibitem{theory-Y-states-chenhuaxing-2016} H. X. Chen, W. Chen, X. Liu, and S. L. Zhu,  The hidden-charm pentaquark and tetraquark states,
 \href{https://www.sciencedirect.com/science/article/pii/S037015731630103X?via%3Dihub}{  Phys.\ Rept.\ }{\bf 639}, 1 (2016).

\bibitem{theory-Y-states-Esposito-2017} A. Esposito, A. Pilloni and A. D. Polosa,  Multiquark resonances,
 \href{https://www.sciencedirect.com/science/article/pii/S037015731630391X?via%3Dihub}{  Phys.\ Rept.\ }{\bf 668}, 1 (2017).

\bibitem{theory-Y-states-Richard-2017} R. F. Lebed, R. E. Mitchell, and E. S.  Swanson,  Heavy-quark QCD exotica, 
 \href{https://linkinghub.elsevier.com/retrieve/pii/S0146641016300734}{ Prog.\ Part.\ Nucl.\ Phys.\ {\bf 93}, 143 (2017)}.

\bibitem{theory-Y-states-Ali-2017} A. Ali, J. S. Lange and S. Stone,  Exotics: Heavy pentaquarks and tetraquarks, 
 \href{https://www.sciencedirect.com/science/article/pii/S0146641017300716?via%3Dihub}{ Prog.\ Part.\ Nucl.\ Phys.\ {\bf 97}, 123 (2017)}.

\bibitem{theory-Y-states-Stephen-2018} S. L. Olsen, T. Skwarnicki, and D. Zieminska, Nonstandard heavy mesons and baryons: Experimental evidence, 
 \href{https://journals.aps.org/rmp/abstract/10.1103/RevModPhys.90.015003}{  Rev.\ Mod.\ Phys.\ {\bf 90}, 015003 (2018)}.

\bibitem{theory-Y-states-guofenghun-2018} F. K. Guo, C. Hanhart, U. G. Mei\ss ner, Q. Wang, Q. Zhao, and B. S. Zou,  Hadronic molecules, 
 \href{https://journals.aps.org/rmp/abstract/10.1103/RevModPhys.90.015004}{  Rev.\ Mod.\ Phys.\ {\bf 90}, 015004 (2018)}.

\bibitem{intro-BaBar-Y4260} B. Aubert {\it et al.} (BaBar Collaboration),  Observation of a broad structure in the $\ppjpsi$ mass spectrum around $4.26$ GeV/$c^2$, 
 \href{https://journals.aps.org/prl/abstract/10.1103/PhysRevLett.95.142001}{Phys.\ Rev.\ Lett.\ {\bf 95}, 142001 (2005)}.

 \bibitem{intro-CLEO-Y4260-1} Q. He {\it et al.} (CLEO Collaboration),  Confirmation of the 
$Y(4260)$ resonance production in initial state radiation, 
 \href{https://journals.aps.org/prd/abstract/10.1103/PhysRevD.74.091104}{ Phys.\ Rev.\ D {\bf 74},  091104(R) (2006)}.

\bibitem{intro-Belle-Y4260} C. Z. Yuan  {\it et al.} (Belle Collaboration),  Measurement of the 
$\ee\to\ppjpsi$ cross section via initial-state radiation at Belle, 
 \href{https://journals.aps.org/prl/abstract/10.1103/PhysRevLett.99.182004}{  Phys.\ Rev.\ Lett.\ {\bf 99}, 182004 (2007)}.

\bibitem{intro-BaBar-Y4260-2012} J. P. Lees  {\it et al.} (BaBar Collaboration),  Study of the reaction $\ee\to\jpsi\pp$ via initial-state radiation at BaBar, 
 \href{https://journals.aps.org/prd/abstract/10.1103/PhysRevD.86.051102}{  Phys.\ Rev.\ D {\bf 86},   051102(R) (2012)}.

\bibitem{intro-Belle-Y4260-2} Z. Q. Liu  {\it et al.} (Belle Collaboration),  Study of  $\ee\to\ppjpsi$  and observation of a charged charmoniumlike state at Belle, 
 \href{https://journals.aps.org/prl/abstract/10.1103/PhysRevLett.110.252002}{  Phys.\ Rev.\ Lett.\ {\bf 110},  252002  (2013)}.

\bibitem{intro-BaBar-Y4360} B. Aubert  {\it et al.} (BaBar Collaboration),  Evidence of a broad structure at an invariant mass of  $4.32$ GeV/$c^2$  in the reaction $\ee\to\pp\psip$ measured at BaBar, 
 \href{https://journals.aps.org/prl/abstract/10.1103/PhysRevLett.98.212001}{   Phys.\ Rev.\ Lett.\ {\bf 98},  212001 (2007)}.

\bibitem{intro-Belle-Y4360-Y4660} X. L. Wang  {\it et al.} (Belle Collaboration),  Observation of two resonant structures in $\ee\to\pp\psip$ via initial-state radiation at Belle, 
 \href{https://journals.aps.org/prl/abstract/10.1103/PhysRevLett.99.142002}{  Phys.\ Rev.\ Lett.\ {\bf 99}, 142002 (2007)}.

\bibitem{intro-BaBar-Y4360-Y4660-2014} J. P. Lees  {\it et al.} (BaBar Collaboration), Study of the reaction $\ee\to\psip\pp$  via initial-state radiation at BaBar, 
   \href{https://journals.aps.org/prd/abstract/10.1103/PhysRevD.89.111103}{  Phys.\ Rev.\ D {\bf 89},  111103(R) (2014)}.

\bibitem{intro-Y4220-bes3-open-charm}M. Ablikim {\it et al.} (BESIII Collaboration),   Evidence of a resonant structure in the $\ee\to\pip\dz\dsm$  cross section between $4.05$ and $4.60$ GeV, 
 \href{https://journals.aps.org/prl/abstract/10.1103/PhysRevLett.122.102002}{ Phys.\ Rev.\ Lett.\ {\bf 122}, 102002 (2019)}.

  \bibitem{intro-Y4360-bes3-open-charm-2022} M. Ablikim {\it et al.} (BESIII Collaboration),  Measurement of $\ee\to\pip\pim\dplus\dm$ cross sections at center-of-mass energies from $4.190$ to $4.946$ GeV, 
 \href{https://link.aps.org/doi/10.1103/PhysRevD.106.052012}{ Phys.\ Rev.\ D\ {\bf 106}, 052012 (2022)}.
 
  \bibitem{Y4630-belle-2020}S. Jia  {\it et al.} (Belle Collaboration),    Evidence for a vector charmoniumlike state in $\ee\to D^{+}_{s}D^{*}_{s2}(2573)^{-}+c.c.$, 
   \href{https://journals.aps.org/prd/pdf/10.1103/PhysRevD.101.091101}{ Phys.\ Rev.\ D {\bf 101},  091101(R)  (2020)}.

\bibitem{Ystate-DDpi-bes3-2023} M. Ablikim {\it et al.} (BESIII Collaboration), Observation of Three Charmoniumlike States with $\jpc=1^{--}$  in $\ee\to\dsz\dsm\pip$ 
 \href{https://journals.aps.org/prl/abstract/10.1103/PhysRevLett.130.121901}{  Phys.\ Rev.\ Lett.\ {\bf 130},   121901 (2023)}.


  \bibitem{kkjpsi-bes3-2022} M. Ablikim {\it et al.} (BESIII Collaboration),  Observation of the $Y(4230)$ and a new structure in $e^+e^- \rightarrow K^+K^-J/\psi$, 
 \href{https://dx.doi.org/10.1088/1674-1137/ac945c}{ Chin.\ Phys.\ C\ {\bf 46}, 111002 (2022)}.  


\bibitem{etapsip} M. Ablikim {\it et al.} (BESIII Collaboration),  Observation of $\ee\to\eta\psip$ at center-of-mass energies from 4.236 to 4.600 GeV, \href{https://doi.org/10.1007/JHEP10(2021)177}{ J. High Energ. Phys. {\bf 2021}, 177 (2021)}.


\bibitem{compass} M. Aghasyan {\it et al.} (COMPASS Collaboration), Search for muoproduction of $X(3872)$ at COMPASS and indication of a
new state  $\tx $, \href{https://doi.org/10.1016/j.physletb.2018.07.008}{Phys. Lett. B {\bf 783}, 334 (2018)}.

\bibitem{pdg} R. Workman {\it et al.} (Particle Data Group),  The Review of particle physics, 
\href{https://academic.oup.com/ptep/article/2020/8/083C01/5891211}{ Prog.\ Theor.\ Exp.\ Phys.\ {\bf 2022}, 083C01 (2022)}.

\bibitem{Ablikim-2009aa} M. Ablikim {\it et al.} (BESIII Collaboration),   Design and construction of the BESIII detector, 
   \href{https://www.sciencedirect.com/science/article/pii/S0168900209023870?via\%3Dihub}{  Nucl.\ Instrum.\ Meth.\  A {\bf 614}, 345 (2010)}.

\bibitem{Yu-IPAC2016-TUYA01}
   C.~H.~Yu {\it et al.},  BEPCII performance and beam dynamics studies on luminosity,  
   \href{http://accelconf.web.cern.ch/ipac2016/doi/JACoW-IPAC2016-TUYA01.html}{ in  Proceedings of IPAC$2016$, Busan, Korea, 2016}.

\bibitem{bes3-energy-measurement}
  M. Ablikim {\it et al.} (BESIII Collaboration),   Measurements of the center-of-mass energies at BESIII via the di-muon process, 
  \href{http://hepnp.ihep.ac.cn/article/doi/10.1088/1674-1137/40/6/063001}{ Chin.\ Phys.\ C {\bf 40}, 063001 (2016)}.

  \bibitem{Ablikim:2019hff}
  M.~Ablikim {\it et al.} [BESIII Collaboration],
  Chin. Phys. C {\bf 44}, 040001 (2020).

\bibitem{EcmsMea}
  J.~Lu, Y.~Xiao, X.~Ji,
  Radiat. Detect. Technol. Methods {\bf 4}, 337–344 (2020).

\bibitem{EventFilter}
  J.~W.~Zhang, L.~H.~Wu, S.~S.~Sun {\it et al.},
  Radiat. Detect. Technol. Methods {\bf 6}, 289–293 (2022).

\bibitem{luminosity-measurement}
M. Ablikim {\it et al.} (BESIII Collaboration),   Precision measurement of the integrated luminosity of the data taken by BESIII at center-of-mass energies between $3.810$ GeV and $4.600$ GeV, 
 \href{http://hepnp.ihep.ac.cn/article/doi/10.1088/1674-1137/39/9/093001}{  Chin.\ Phys.\ C {\bf 39}, 093001 (2015)}.

  \bibitem{luminosity-measurement-2}
M. Ablikim {\it et al.} (BESIII Collaboration),   Luminosity measurements for the R scan experiment at BESIII, 
  \href{http://hepnp.ihep.ac.cn/en/article/doi/10.1088/1674-1137/41/6/063001}{ Chin.\ Phys.\ C {\bf 41}, 063001 (2017).}.

\bibitem{etof}
 X.~Li {\it et al.},  Study of MRPC technology for BESIII endcap-TOF upgrade, 
  \href{https://link.springer.com/article/10.1007/s41605-017-0014-2}{  Rad.\ Det.\ Tech.\ Meth.\ {\bf 1}, 13 (2017)}.

  \bibitem{etof-2}
 Y.~X.~Guo {\it et al.},  The study of time calibration for upgraded end cap TOF of BESIII, 
  \href{https://link.springer.com/article/10.1007\%2Fs41605-017-0012-4}{  Rad.\ Det.\ Tech.\ Meth.\ {\bf 1}, 15 (2017)}.

   \bibitem{etof-3}
 P.~Cao {\it et al.},  Design and construction of the new BESIII endcap Time-of-Flight system with MRPC technology, 
  \href{https://www.sciencedirect.com/science/article/pii/S0168900219314068}{  Nucl.\ Instrum.\ Meth.\ A {\bf 953}, 163053 (2020)}.

\bibitem{geant4}S. Agostinelli  {\it et al.}  (GEANT4 Collaboration),  Geant$4$---a simulation toolkit, 
 \href{https://www.sciencedirect.com/science/article/pii/S0168900203013688?via\%3Dihub}{ Nucl.\ Instrum.\ Meth.\ A {\bf 506}, 250 (2003)}.

\bibitem{ref-kkmc}
  S.~Jadach, B.~F.~L.~Ward, and Z.~Was,   The precision Monte Carlo event generator KK  for two-fermion final states in  $\ee$  collisions, 
  \href{https://www.sciencedirect.com/science/article/pii/S0010465500000485?via\%3Dihub}{  Comput.\ Phys.\ Commun.\ {\bf 130}, 260 (2000)}.

  \bibitem{ref-kkmc-2}
  S.~Jadach, B.~F.~L.~Ward, and Z.~Was,  Coherent exclusive exponentiation for precision Monte Carlo calculations, 
  \href{https://journals.aps.org/prd/abstract/10.1103/PhysRevD.63.113009}{ Phys.\ Rev.\ D  {\bf 63}, 113009 (2001)}.

  \bibitem{ref-evtgen}
  D.~J.~Lange,  The EvtGen particle decay simulation package, 
  \href{https://www.sciencedirect.com/science/article/pii/S0168900201000894?via\%3Dihub}{  Nucl.\ Instrum.\ Meth.\ A {\bf 462}, 152 (2001)}.

\bibitem{ref-evtgen-2}
      R.~G.~Ping,  Event generators at BESIII, 
  \href{https://iopscience.iop.org/article/10.1088/1674-1137/32/8/001}{ Chin.\ Phys.\ C {\bf 32}, 599 (2008)}.

\bibitem{photos}
  E.~Richter-Was,  QED bremsstrahlung in semileptonic B and leptonic $\tau$ decays, 
   \href{https://www.sciencedirect.com/science/article/abs/pii/037026939390062M?via\%3Dihub}{  Phys.\ Lett.\ B {\bf 303},  163 (1993)}.

\bibitem{ref:lundcharm}
 J.\ C.\ Chen, G.\ S.\ Huang, X.\ R.\ Qi, D.\ H.\ Zhang, and Y.\ S.\ Zhu,  Event generator for $\jpsi$ and $\psip$ decay, 
  \href{https://journals.aps.org/prd/abstract/10.1103/PhysRevD.62.034003}{ Phys.\ Rev.\ D {\bf 62},  034003 (2000)}.

\bibitem{ref:lundcharm2}
 R.\ L.\ Yang, R.\ G.\
Ping, and H.\ Chen,  Tuning and validation of the Lundcharm model with $\jpsi$ decays , \href{https://iopscience.iop.org/article/10.1088/0256-307X/31/6/061301}{ Chin.\ Phys.\ Lett.\  {\bf 31}, 061301 (2014)}.
 
\bibitem{vacuum-polarization-factor}
  S.~Actis {\it et al.} (Working Group on Radiative Corrections and Monte Carlo Generators for Low Energies),  Quest for precision in hadronic cross sections at low energy: Monte Carlo tools vs.\ experimental data, 
   \href{https://link.springer.com/article/10.1140\%2Fepjc\%2Fs10052-010-1251-4}{  Eur.\ Phys.\ J.\ C {\bf 66}, 585 (2010)}.

\bibitem{bes3-Y4230-omegachic0} M. Ablikim {\it et al.} (BESIII Collaboration),  Cross section measurements of $\ee\to\omega\chi_{c0}$  from $\sqrt{s}=4.178$ to $4.278$ GeV, 
 \href{https://journals.aps.org/prd/abstract/10.1103/PhysRevD.99.091103}{ Phys.\ Rev.\ D {\bf 99}, 091103(R) (2019)}.

\bibitem{cross-section-pppsip} M. Ablikim {\it et al.} (BESIII Collaboration),   Measurement of $\ee\to\pp\psi(3686)$ from $4.008$ to $4.600$ GeV and observation of a charged structure in the $\pi^{\pm}\psi(3686)$ mass spectrum,  
 \href{https://journals.aps.org/prd/abstract/10.1103/PhysRevD.96.032004}{  Phys.\ Rev.\ D {\bf 96},  032004  (2017)}.

\bibitem{cross-section-p0p0psip} M. Ablikim {\it et al.} (BESIII Collaboration),  Measurement of $\ee\to\piz\piz\psi(3686)$ at $\sqrt{s}$ from $4.009$ to $4.600$ GeV and observation of a neutral charmoniumlike structure, 
  \href{https://journals.aps.org/prd/abstract/10.1103/PhysRevD.97.052001}{ Phys.\ Rev.\ D {\bf 97},  052001 (2018)}.

\bibitem{cross-section-omegachic012} M. Ablikim {\it et al.} (BESIII Collaboration),  Study of $\ee\to\omega\chi_{cJ}$ at center of mass energies from $4.21$ to $4.42$ GeV, 
 \href{https://journals.aps.org/prl/abstract/10.1103/PhysRevLett.114.092003}{  Phys.\ Rev.\ Lett.\ {\bf 114}, 092003 (2015)}.

  \bibitem{cross-section-omegachic012-2} M. Ablikim {\it et al.} (BESIII Collaboration), 
 Observation of $\ee\to\omega\chi_{c1,2}$ near $\sqrt{s}=4.42$ and $4.6$ GeV, 
  \href{https://journals.aps.org/prd/abstract/10.1103/PhysRevD.93.011102}{ Phys.\ Rev.\ D {\bf 93}, 011102(R) (2016)}.

\bibitem{cross-section-gammax3872}
M. Ablikim {\it et al.} (BESIII Collaboration),   Study of $\ee\to\gamma\omega\jpsi$ and observation of $X(3872)\to\omega\jpsi$, 
  \href{https://journals.aps.org/prl/abstract/10.1103/PhysRevLett.122.232002}{ Phys.\ Rev.\ Lett.\ {\bf 122}, 232002 (2019)}.

\bibitem{cross-section-phichic12} M. Ablikim {\it et al.} (BESIII Collaboration),   Observation of $\ee\to\phi\chi_{c1}$ and $\phi\chi_{c2}$  at $\sqrt{s}=4.600$ GeV,
  \href{https://journals.aps.org/prd/abstract/10.1103/PhysRevD.97.032008}{ Phys.\ Rev.\ D {\bf 97}, 032008 (2018)}.

\bibitem{isr-calculate2}
E. A. Kuraev and V. S. Fadin,  Yad.\  Fiz.\ {\bf 41},  733  (1985)  [Sov.\ J.\  Nucl.\ Phys.\ {\bf 41}, 466 (1985)].

\bibitem{feldman-cousins-method} G. J.~Feldman and R. D. ~Cousins.  Unified approach to the classical statistical analysis of small signals, 
  \href{https://journals.aps.org/prd/abstract/10.1103/PhysRevD.57.3873}{ Phys.\ Rev.\ D {\bf 57}, 3873 (1998)}.

\bibitem{pole-method} J. Conrad,  A program for confidence interval calculations for a Poisson process with background including systematic uncertainties: POLE $1.0$, 
 \href{https://www.sciencedirect.com/science/article/pii/S0010465504000037}{ Comput.\ Phys.\ Commun.\ {\bf 2}, 158 (2004)}.

\bibitem{intro-bes3-etapjpsi} M. Ablikim {\it et al.} (BESIII Collaboration),  Observation of 
$\ee\to\eta'\jpsi$  at center-of-mass energies between $4.189$ and $4.600$ GeV, 
 \href{https://journals.aps.org/prd/abstract/10.1103/PhysRevD.94.032009}{  Phys.\ Rev.\ D {\bf 94}, 032009 (2016)}.

\bibitem{photoneffi}M. Ablikim {\it et al.} (BESIII Collaboration),   Branching fraction measurements of $\chi_{c0}$  and $\chi_{c2}$ to $\piz\piz$ and $\eta\eta$, 
 \href{https://journals.aps.org/prd/abstract/10.1103/PhysRevD.81.052005}{  Phys.\ Rev.\  D {\bf 81}, 052005 (2010)}.

\bibitem{bes3-kinematicfit-eff}
M. Ablikim {\it et al.} (BESIII Collaboration),  Search for hadronic transition $\chi_{cJ}\to\eta_{c}\pp$ and observation of $\chi_{cJ}\to K\bar{K}\pi\pi\pi$, 
 \href{https://journals.aps.org/prd/abstract/10.1103/PhysRevD.87.012002}{ Phys.\ Rev.\ D {\bf 87}, 012002 (2013)}.

\end{thebibliography}
\end{document}